\documentclass[10pt,twocolumn,letterpaper]{article}

\usepackage[pagenumbers]{cvpr} 

\usepackage{graphicx}
\usepackage{amsmath}
\usepackage{amssymb}
\usepackage{booktabs}
\usepackage{setspace} 
\usepackage{multirow}
\usepackage{arydshln}
\usepackage{xcolor}
\usepackage{float}
\usepackage{tikz}
\usepackage{pgfplots}
\usepackage{subcaption} 
\pgfplotsset{compat=1.18}

\usepackage[pagebackref,breaklinks,colorlinks]{hyperref}

\usepackage[capitalize]{cleveref}
\crefname{section}{Sec.}{Secs.}
\Crefname{section}{Section}{Sections}
\Crefname{table}{Table}{Tables}
\crefname{table}{Tab.}{Tabs.}

\def\cvprPaperID{12800} 
\def\confName{CVPR}
\def\confYear{2023}

\begin{document}

\title{CFAT: Unleashing Triangular Windows for Image Super-resolution}

\author{Abhisek Ray\\
Indian Institute of Technology Patna\\
{\tt\small rayabhisek0610@gmail.com}
\and
Gaurav Kumar\\
Indian Institute of Technology Patna\\
{\tt\small gaurav19.iitp@gmail.com}
\and
Maheshkumar H. Kolekar\\
Indian Institute of Technology Patna\\
{\tt\small mahesh@iitp.ac.in}\\
\href{https://github.com/rayabhisek123/CFAT}{https://github.com/rayabhisek123/CFAT}
}

\maketitle

\begin{abstract}   
	\label{sec:abstract}
	Transformer-based models have revolutionized the field of image super-resolution (SR)  by harnessing their inherent ability to capture complex contextual features. The overlapping rectangular shifted window technique used in transformer architecture nowadays is a common practice in super-resolution models to improve the quality and robustness of image upscaling. However, it suffers from distortion at the boundaries and has limited unique shifting modes. To overcome these weaknesses, we propose a non-overlapping triangular window technique that synchronously works with the rectangular one to mitigate boundary-level distortion and allows the model to access more unique sifting modes. In this paper, we propose a Composite Fusion Attention Transformer (CFAT) that incorporates triangular-rectangular window-based local attention with a channel-based global attention technique in image super-resolution. As a result, CFAT enables attention mechanisms to be activated on more image pixels and captures long-range, multi-scale features to improve SR performance. The extensive experimental results and ablation study demonstrate the effectiveness of CFAT in the SR domain. Our proposed model shows a significant 0.7 dB performance improvement over other state-of-the-art SR architectures.
\end{abstract}
\vspace*{-5mm}

\section{Introduction}
\label{sec:Introduction}
\vspace*{-1mm}
Efficient transmission or storage of images in band- or memory-limited systems heavily relies on diverse image compression techniques. However, in the real world, these image compression phenomenons are lossy. It is challenging to retrieve the original visual quality from its compressed correspondent. Image super-resolution (SR) \cite{dong2014learning}, a reverse engineering approach in contrast to compression, aims to recover high-resolution (HR) images from their low-resolution (LR) counterparts. Despite extensive and consistent research efforts, SR models still struggle to eliminate the visual artifacts in recovered images entirely. In early 2014, the evolution of deep neural networks enhanced SR capabilities with CNN as a mainframe skeleton \cite{dong2014learning}. These approaches \cite{zhang2018residual, zhang2018image, kong2021classsr, dai2019second} frequently come with substantial computational and memory requirements regardless of their effectiveness. The transformer-based techniques have stood out as the most successful SR approaches among many emerging models, showcasing state-of-the-art performances in the recent era.

The surge in popularity of transformer-based architectures is attributed to their natural proficiency in leveraging long-range dependencies among image features. Leading approaches such as SwinIR\cite{liang2021swinir}, Swin2SR\cite{conde2022swin2sr}, ESRT\cite{lu2022transformer}, HST\cite{li2022hst}, HAT\cite{chen2023activating} and ART\cite{zhang2022accurate} advocate the above benefit. They predominantly rely on a hierarchical attention mechanism, ensuring the reconstruction of HR outputs from their LR counterparts with enhanced precision. 

\begin{figure}[t]
	\centering
	\captionsetup{justification=centering}
	\includegraphics[width=0.95\linewidth]{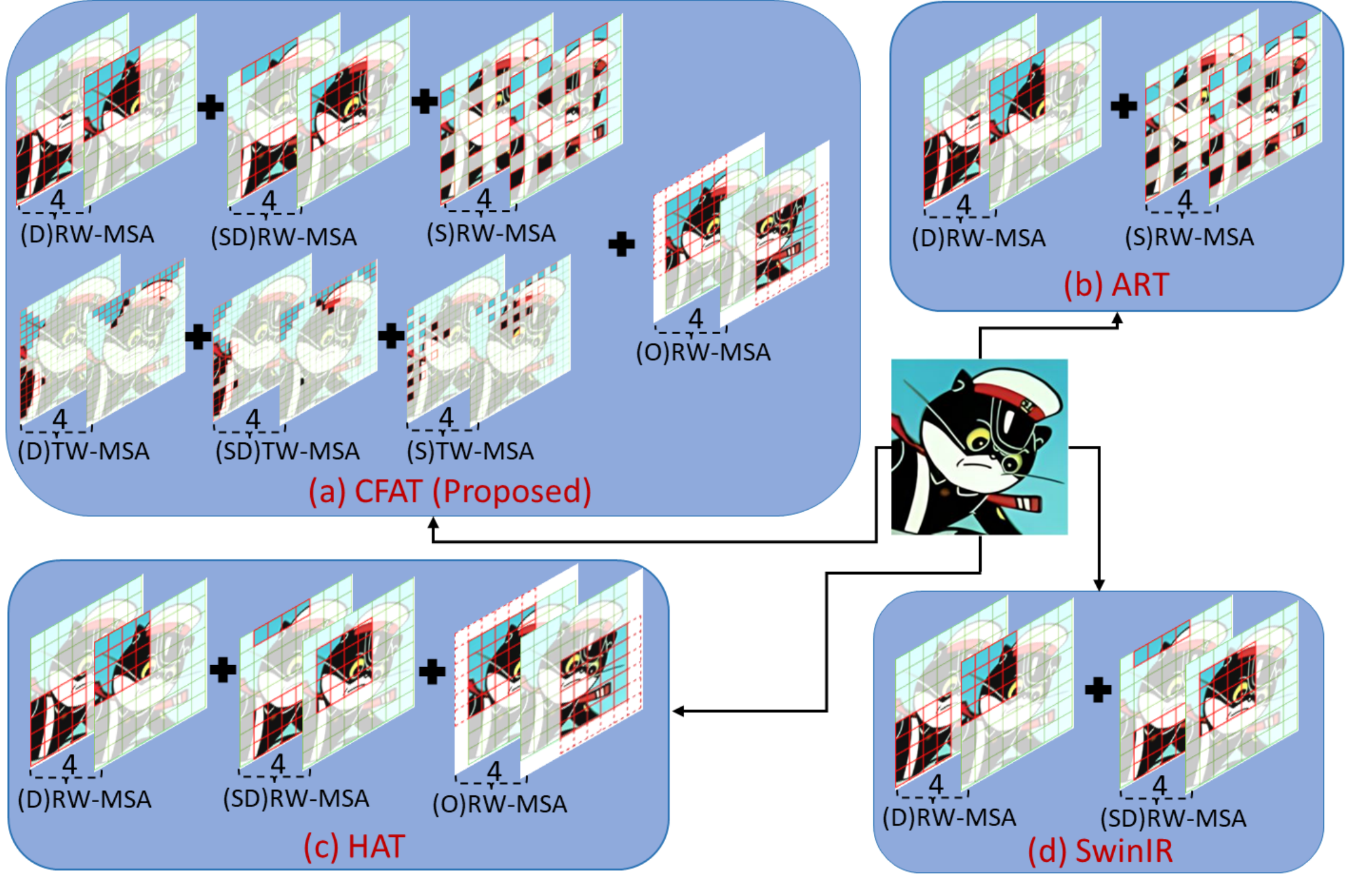}
	\vspace*{-3mm}
	\caption{Proposed CFAT vs other SOTA models. RW/TW: Rectangular/Triangular Window, MSA: Multi-Head Attention, (D): Dense, (SD): Shifted Dense, (S): Sparse, (O): Overlapping}
	\label{fig:Novelty}
	\vspace*{-7mm}
\end{figure}
SwinIR \cite{liu2021swin} pioneered the application of the shifted window technique in the domain of image restoration \cite{liang2021swinir}, focusing mainly on image super-resolution. Swin2SR \cite{conde2022swin2sr} redefined the SwinIR to enhance the performance after effectively tackling data and training-related challenges. However, both \cite{liang2021swinir} and \cite{conde2022swin2sr} consistently concentrate exclusively on specific image patches for self-attention. The dilated window approach introduced in ART \cite{zhang2022accurate} aims not only to resolve the fixed window problem as mentioned earlier but also to alleviate computational load while concurrently boosting super-resolution performance. Nevertheless, the quick internal switching between dense and sparse attention layers, each tailored to specific receptive fields, imposes constraints on overall performance. HAT \cite{chen2022activating} outperforms other prominent super-resolution architectures by a significant 1 dB margin, achieved through integrating global channel attention with local window-based self-attention. The incorporation of the overlapping shifted window technique effectively addresses concerns related to repetitive image patches, while the introduction of channel attention substantially enhances super-resolution performance. Drawing inspiration from the CNN-transformer model, ACT \cite{yoo2023enriched} strategically positions both CNN and transformer within two distinct streams to concurrently leverage global and local features. The harmonious, complementary features on both branches interchange relatable information with one another through a channel split fusion block based on cross-scale token attention (CSTA). The above-discussed state-of-the-art transformer-based methods use a rectangular shifted window-based self-attention mechanism. The repetitive use of rectangular window attention at each layer is susceptible to causing boundary-level distortion, as the participation of neighbors for boundary pixels is limited. In addition to this, the symmetrical structure of the rectangular window restricts the range of unique shifting modes available for covering the entire image patches. The presence of boundary-level distortion and a limited range of shifting modes in the shifted rectangular window technique could potentially hinder its effectiveness, extending beyond super-resolution to other computer vision tasks. 

To tackle the challenges mentioned earlier, we propose a pioneering triangular window technique. This novel window technique helps boost the performance for super-resolution tasks and all computer vision applications where the rectangular window technique is predominantly used. This paper seamlessly incorporates the proposed triangular window with the conventional rectangular window in a transformer framework called Composite Fusion Attention Transformer (CFAT). Integrating a channel-based global attention unit with the above triangular-rectangular window-based local self-attention further improves the performance of single-image SR. The above attention pair belongs to the non-overlapping attention category and effectively leverages local and global image features. To exploit the overlapping image features, we place an overlapping cross-fusion attention block (OCFAB) at the end of each unit block called the Dense Window Attention Block (DWAB). The proposed model, CFAT, offers several unique advantages: (i) The combined window-based self-attention technique eliminates the boundary level distortion issue that arises during traditional rectangular window-based self-attention. (ii) Embracing a triangular window offers a broader range of shifting modes, thereby expanding spatial features and enhancing overall performance. (iii) OCFAB leverages overlapping spatial image features, resulting in a further enhancement of performance.

The proposed CFAT network delivers superior SR results across multiple benchmark datasets, notably outperforming recent transformer-based SR techniques \cite{li2022hst, liu2021swin}. The novelty of our paper in comparison to other state-of-the-art methods like SwinIR \cite{liu2021swin}, ART \cite{zhang2022accurate}, and HAT \cite{chen2022activating} is outlined in Fig. \ref{fig:Novelty}. Our contributions in this paper can be summarized as:
\vspace*{-2mm}
\begin{enumerate}
	\item We are the first to introduce the triangular windowing mechanism in the computer vision task. We smoothly integrate it with traditional rectangular windows to employ non-overlapping self-attention in single-image SR. This combination phases out the constraints associated with the conventional rectangular window approaches like boundary-level distortion and restricted unique shifting modes.
	\vspace*{-2.5mm}
	\item This window mechanism is beneficial not only in super-resolution tasks but also in various other computer vision applications that implement the rectangular window technique in their mainframe.
	\vspace*{-2.5mm}
	\item We propose a CFAT that exploits local and global image features through overlapping and non-overlapping spatial and channel attention, respectively.
	\vspace*{-2.5mm}
	\item Through comprehensive evaluations on multiple benchmark SR datasets, our method has delivered superior performance compared to other state-of-the-art methods.
\end{enumerate}
\vspace*{-4mm}
\begin{figure*}[t]
	\centering
	\captionsetup{justification=centering}
	\includegraphics[width=0.81\linewidth]{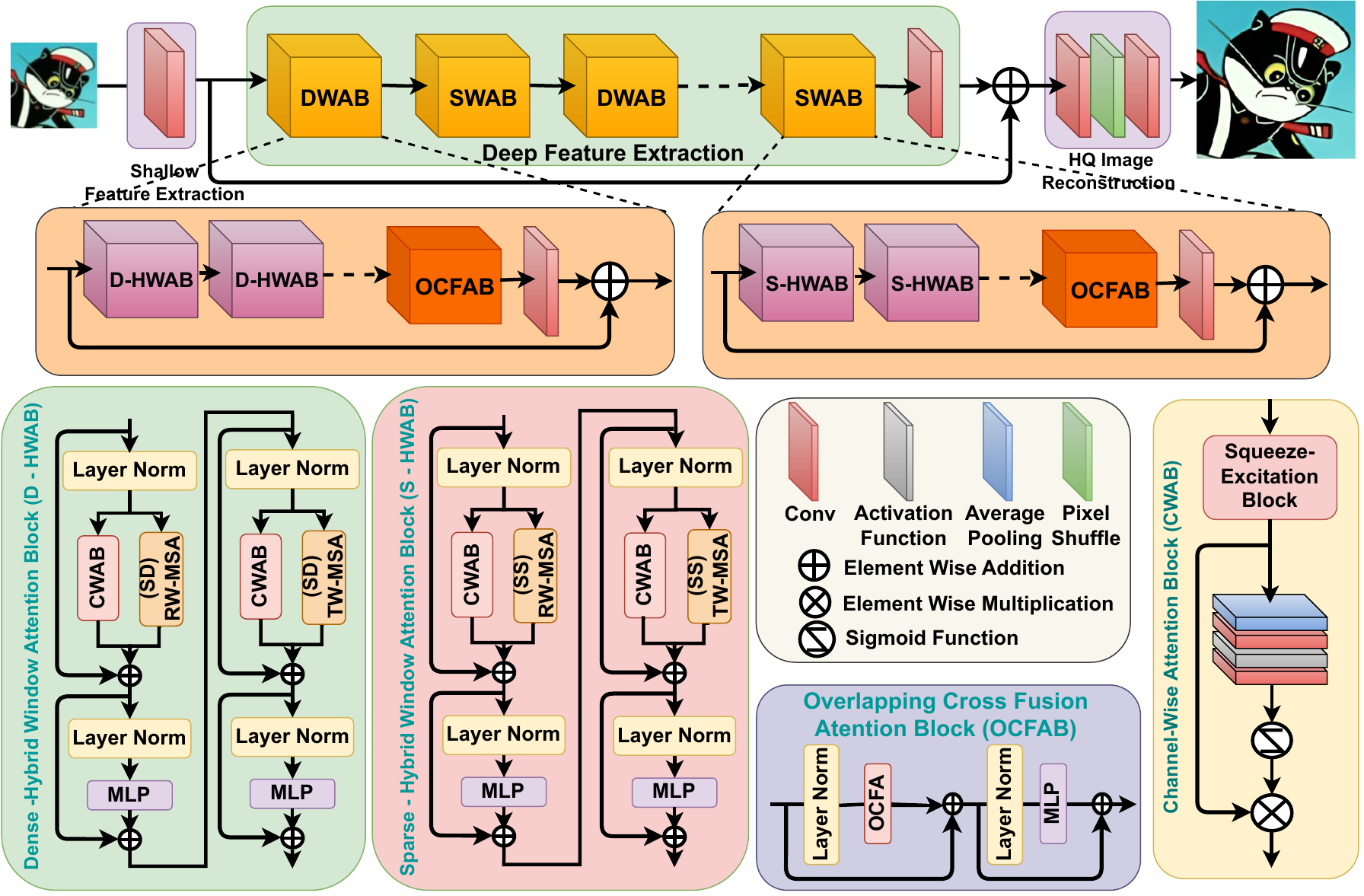}
	\vspace*{-3mm}
	\caption{The overall architecture of CFAT with all internal units.}
	\label{fig:model_arch}
	\vspace*{-6mm}
\end{figure*}

\section{Related Works}
\label{sec:Related_Works}

\subsection{CNN Based Super-Resolution}
\vspace*{-1mm}
Initially, most single-image super-resolution (SISR) approaches relied heavily on convolutional neural networks (CNNs) due to their remarkable spatial feature extraction capability over traditional machine learning approaches. SRCNN \cite{dong2015image} stands as a pioneering effort, leveraging a shallow three-layer CNN architecture to map low-resolution (LR) images to their high-resolution (HR) counterparts. However, a sharp decline in performance is observed while increasing the number of CNN layers intended for extracting more complex features. EDSR \cite{lim2017enhanced} is an SR architecture based on the Residual Network (ResNet) framework, which mitigates this limitation by including shortcut connections. Introducing channel attention mechanisms in super-resolution, RCAN \cite{zhang2018image} presents a unique residual-in-residual (RIR) design. RDN \cite{zhang2018residual} utilizes a residual dense block (RDB) that efficiently extracts ample local features through densely interconnected CNN layers. SAN \cite{dai2019second} is built upon a second-order attention network that takes SR performance to a new height. LatticeNet \cite{luo2020latticenet} incorporates the Fast Fourier Transform (FFT) technique to construct a lattice design, enabling lightweight super-resolution. \cite{dong2016accelerating, guo2020closed, kim2016accurate, ledig2017photo, lim2017enhanced, niu2020single}, and \cite{hui2018fast} are other prominent CNN-based SR models that demonstrate notable advancement in performances over their predecessors. These models incorporate advanced CNN architectures, such as residual blocks \cite{zhang2018image, lim2017enhanced}, dense blocks \cite{zhang2018residual}, and attention mechanisms \cite{niu2020single, guo2020closed}, to enhance feature generalization, ultimately leading to superior outcomes. While CNN-based methods have seen significant success in SISR, they do come with limitations. Specifically, they can only extract local features from images, and they often entail high computational costs for both training and deployment. 
\vspace*{-2mm}
\subsection{Vision Transformer (ViT) Based SR}
\vspace*{-1mm}
The transformer architecture \cite{vaswani2017attention} operates on the principle of self-attention, a technique that generalizes features in terms of a global spatial perspective and can capture long-range dependencies between them. ViT, introduced in \cite{dosovitskiy2020image}, is evident as the first work to substitute standard convolution with the transformers in high-level vision tasks. Following this, in many computer vision applications, including image recognition \cite{meng2022adavit, li2021bringing}, object detection \cite{carion2020end, chu2021twins}, image segmentation \cite{cao2022swin, huang2022glance}, and image SR \cite{zhu2023attention, qiu2023dual, yu2023dipnet, luo2022deep}, transformer-based architecture replaces conventional CNN-based models to elevate their performances. Building on transformer models, SwinIR\cite{liang2021swinir} employs the Swin transformer with window-based self-attention for image restoration. Its improved version, Swin2SR \cite{conde2022swin2sr}, utilizes the SwinV2 transformer specifically for compressed image super-resolution and restoration tasks. Meanwhile, ESRT \cite{lu2022transformer} introduces efficient transformers that achieve competitive results at a reduced computational cost. ART \cite{zhang2022accurate} incorporates both sparse and dense attention modules within a single network that aim to refine image restoration results. In a recent advancement, \cite{chen2023activating} introduced a transformer architecture, HAT, which combines channel attention with overlapping and non-overlapping self-attention mechanisms to activate more pixels, thereby generalizing both global and local spatial features. Although the Vision Transformer has shown its superiority in modeling long-range dependencies, there are still many works that demonstrate that the convolution can help the transformer to achieve better visual representation \cite{li2023uniformer, wu2021cvt, xiao2021early}. Drawing inspiration from the CNN-transformer model, ACT \cite{yoo2023enriched} strategically places both CNN and transformer within two distinct streams to concurrently leverage global and local features. The above SR models adopt a shifted rectangular window technique to limit their spatial features for effective localized self-attention. In contrast, these models experience boundary-level distortion and encounter limitations in shifting modes when exclusively relying on the rectangular window technique. In this paper, we propose a pioneering triangular window technique that works alongside a rectangular one to wipe out these drawbacks in SISR.  
\vspace*{-3mm}

\section{Proposed Method}
\label{sec:Proposed_Method}
\vspace*{-2mm}
\subsection{Overall Architecture}
\label{sec:Overall_Architecture}
\vspace*{-2mm}
As presented in Fig.\ref{fig:model_arch}, the entire network can be split into three segments: the head, body, and tail. The head module is responsible for shallow feature extraction, the body extracts the deep features, and the tail module reconstructs the HR images from LR counterparts at the output stage. The in-depth depiction of the above three is as follows: 
\vspace*{-2mm}
\subsection{Head Module - Shallow Feature Extractor:}
\label{sec:Head_Module}
\vspace*{-2mm}
We feed the input image, $I_{LR} \in \mathbb{R}^{H \times W \times C_{in}}$ to a feature extractor to get shallow output $F_{sh} \in \mathbb{R}^{H \times W \times C}$, where $H$, $W$, $C_{in}$, and $C$ are the height, width, input channel count and output channel count of the respective image. This extractor is a single convolution layer of kernel size $3 \times 3$.
\vspace*{-3mm}
\begin{equation}
	F_{sh} = f_{H}(I_{LR})
	\vspace*{-2mm}
\end{equation}
\subsection{Body Module - Deep Feature Extractor:}
\label{sec:Body_Module}
The alternative units of Dense Window Attention Blocks (DWAB) and Sparse Window Attention Blocks (SWAB) are ultimately combined with a convolution layer to form the deep extractor module. The shallow output from the head module, $F_{sh} \in \mathbb{R}^{H \times W \times C}$, is passed through this deep extractor to yield deep features $F_{dp} \in \mathbb{R}^{H \times W \times C}$. Each unit of DWAB and SWAB is comprised of both rectangular and our proposed triangular window attention units in dense and sparse configurations, respectively.
\vspace*{-3mm}
\subsubsection{Dense Window Attention Blocks (DWAB)}
\label{sec:DWAB}
Each unit of DWAB implements both non-overlapping and overlapping dense window attention that helps to integrate deep and diverse features into a single network. Our design incorporates multiple units of non-overlapping attention units, called Dense-Hybrid Window Attention Block (D-HWAB), initially to get deep features and place an overlapping attention unit, called Overlapping Cross Fusion Attention Block (OCFAB), at the end to get diverse features as well. A $3 \times 3$ convolution layer at the end makes the features richer and more effective, as cited in  \cite{wu2021cvt, xiao2021early, li2023uniformer}. Mathematically, the DWAB can be represented as
\vspace*{-2mm}
\begin{equation}
	F_{dp} = f_{conv}(f_{op}(f^{n}_{nop}...(f^{2}_{nop}(f^{1}_{nop}(F_{sh}))))) + F_{sh},
	\vspace*{-2mm}
\end{equation}
where, $f^{x}_{nop}$, $f_{op}$, and $f_{conv}$ stands for non-overlapping, overlapping and convolution operations, respectively.
\vspace*{-4mm}
\paragraph{Dense-Hybrid Window Attention Block (D-HWAB):}
The underlying module advocates dense attention, alternatively combining shifted rectangular and novel triangular transformer units. The Multi-head Self Attention (MSA) carried out in both units is called Shifted-Dense Rectangular Window MSA ((SD)RW-MSA) and Shifted-Dense Triangular Window MSA ((SD)TW-MSA). Both of them execute non-overlapping self-attention. The combination of channel attention features from Channel Wise Attention Block (CWAB) and spatial attention features from (SD)RW-MSA and (SD)TW-MSA in both transformer units, as shown in Fig.\ref{fig:model_arch}, elevates the overall performance. The overall computation can be represented as
\vspace*{-2mm}
\begin{equation}
	F_{DA} = f^{x}_{nop}(F_{sh}) = f^{n}_{tri}f^{n}_{rect}...(f^{1}_{tri}(f^{1}_{rect}(F_{sh})))),
	\vspace*{-2mm}
\end{equation}
where, $F_{DA}$, $f^{x}_{rect}$, and $f^{x}_{tri}$ are dense attention features, rectangular and triangular window dense attention, respectively. The output feature from rectangular transformer unit can be expressed as
\vspace*{-3mm}
\begin{equation}
	\label{eq:rect}
	F_{int} = f^{rect}_{MSA}(f^1_{LN}(F_{in})) + \alpha f_{CA}(f^1_{LN}(F_{in})) + F_{in},
	\vspace*{-3mm}
\end{equation}
\vspace*{-2mm}
\begin{equation}
	F_{out} = f_{MLP}(f^2_{LN}(F_{int})) + F_{int},
	\vspace*{-2mm}
\end{equation}
where, $F_{in}$, $F_{int}$, and $F_{out}$ are input, intermediate and output features respectively. $f^{rect}_{MSA}$, $f^1_{LN}$, $f^2_{LN}$, and $f_{CA}$ are the feature transformation in (SD)RW-MSA, first LayerNorm, second LayerNorm and CWAB respectively. Similarly, the output feature from triangular transformer unit can be expressed as
\vspace*{-2mm}
\begin{equation}
	\label{eq:tri}
	F_{int} = f^{tri}_{MSA}(f^1_{LN}(F_{in})) + \beta f_{CA}(f^1_{LN}(F_{in})) + F_{in},
	\vspace*{-2mm}
\end{equation}
\begin{equation}
	F_{out} = f_{MLP}(f^2_{LN}(F_{int})) + F_{int},
	\vspace*{-2mm}
\end{equation}
where, $f^{tri}_{MSA}$ is the feature transformation of (SD)TW-MSA. The $\alpha$ in equation \ref{eq:rect} and $\beta$ in equation \ref{eq:tri} are two hyper-parameter to limit the dominancy of channel attention in $F_{out}$ of rectangular and triangular transformer units respectively.
\vspace*{-4mm}
\paragraph{Overlaping Cross Fusion Attention Block (OCFAB):}
This block overlaps the features across neighboring windows and establishes cross-attention between them to improve performance further. The underlying attention is called Overlapping Cross Fusion Attention (OCFA). The OCFAB is implemented by a sliding window approach called unfolding operation, using a kernel of size $R_0$, a stride of $R$, and padding of $ \frac{kR}{2}$ zeros to make the sizes compatible for overlapping. The input feature of dimension $H \times W \times C$ are divided into \( \frac{HW}{R^2} \) overlapping windows of size \( R_0 \). Here, \( R_0 \) is determined by
\vspace*{-2mm}
\begin{equation}
	R_0 = (1+k)R,
	\vspace*{-2mm}
\end{equation} 
where '$k$' is a constant regulating overlapping ratio and $R$ represents the size of window before overlapping. During MSA, the query is calculated by a linear layer, whereas key and value are produced by a linear layer followed by an unfolding operation. The OCFAB block is present at the end of both DWAB and SWAB, as shown in Fig. \ref{fig:model_arch}.
\vspace*{-3mm}
\paragraph{Channe-Wise Attention Block (CWAB):}
Most studies show the practice of using standard convolution while executing squeeze-excitation channel attention \cite{chen2023activating}. However, we use depthwise-pointwise convolution that helps to reduce the squeeze factor without increasing the computational burden. Therefore, we design CWAB with a GELU activation function sandwiched between two depthwise-pointwise convolutional layers followed by a channel-wise attention module at the end.

\vspace*{-5mm}
\subsubsection{Sparse Window Attention Block (SWAB)}
\vspace*{-2mm}
Like DWAB, SWAB promotes non-overlapping and overlapping window attention on sparse windows instead of dense ones. This block also includes four operations as shown in Fig.\ref{fig:model_arch}: (i) non-overlapping rectangular sparse window attention, (ii) non-overlapping triangular sparse window attention, (iii) overlapping cross-fusion attention, and (iv) convolution operation. The latter two are the same as discussed in Sec. \ref{sec:DWAB}, but the former two are a little different in terms of rectangular and triangular window formation compared to DWAB. First, we prepare sparse windows from the input of dimension $H \times W \times C$ by maintaining a regular gap, called interval size $I$, between consecutive input features. After that, we follow the standard rectangular and triangular window techniques to implement respective attention. Sec. \ref{fig:computational_cost} provides more details about triangular sparse attention. We follow a similar procedure in implementing rectangular sparse attention. 
\vspace*{-4mm}
\subsubsection{Tail Module - HQ Image Reconstruction}
\vspace*{-1mm}
In the final stage, we obtain the upscaled input signal ($I_SR$) by passing the deep features ($F_{dp}$) from body through a pixel-shuffle layer sandwiched between a pair of convolution layers. This is represented by ${I}_{SR} = f_{T}(F_{dp})$.
\vspace*{-3mm}

\section{Triangular Window Technique}
\label{fig:computational_cost}
\begin{figure}[t]
	\centering
	\captionsetup{justification=centering}
	\includegraphics[width=0.6\linewidth]{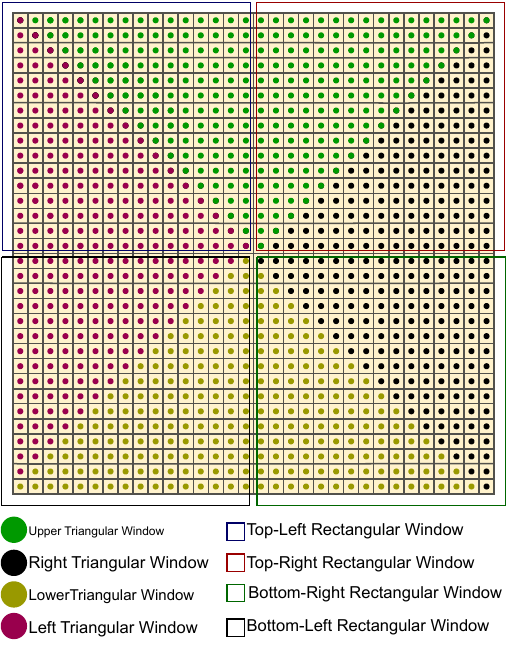}
	\vspace*{-3mm}
	\caption{A rectangular and triangular patch in $32 \times 32$ window.}
	\label{fig:rect_tri}
	\vspace*{-5mm}
\end{figure}

\vspace*{-1mm}
\paragraph{Design:}  Considering an input image $I_{in}$ of dimension $H \times W \times C$, we first divide this input into $\frac{HW}{M^2}$ rectangular windows of size $M \times M$. Then, each rectangular patch is again split into four triangular windows as displayed in Fig.\ref{fig:rect_tri}. In this figure, we take $M$ as 32, which provides four triangular masks of linear dimension $16*16$. These four triangular windows are called upper, right, lower, and left triangular windows. The corresponding rectangular windows are also shown in this figure.
\vspace*{-4mm}
\paragraph{Advantages over Rectangular Window:} In computer vision, the attributes of a pixel depend on itself and its neighbor as well. As a result, the edge pixels of the rectangular window are not as effectively explored as the interior pixels during self-attention, leading to distortions at the boundaries. The distortion will become so pronounced that it will ultimately deteriorate model performance. The above problem can be prominently seen in both rectangular and triangular windows while implemented alone. Therefore, we combine the rectangular window-based MSA with our proposed triangular one in series. This alternative configuration of rectangular and triangular transformers mitigates the edge-level distortion of one another.

\begin{figure}[t]
	\centering
	\captionsetup{justification=centering}
	\includegraphics[width=0.95\linewidth]{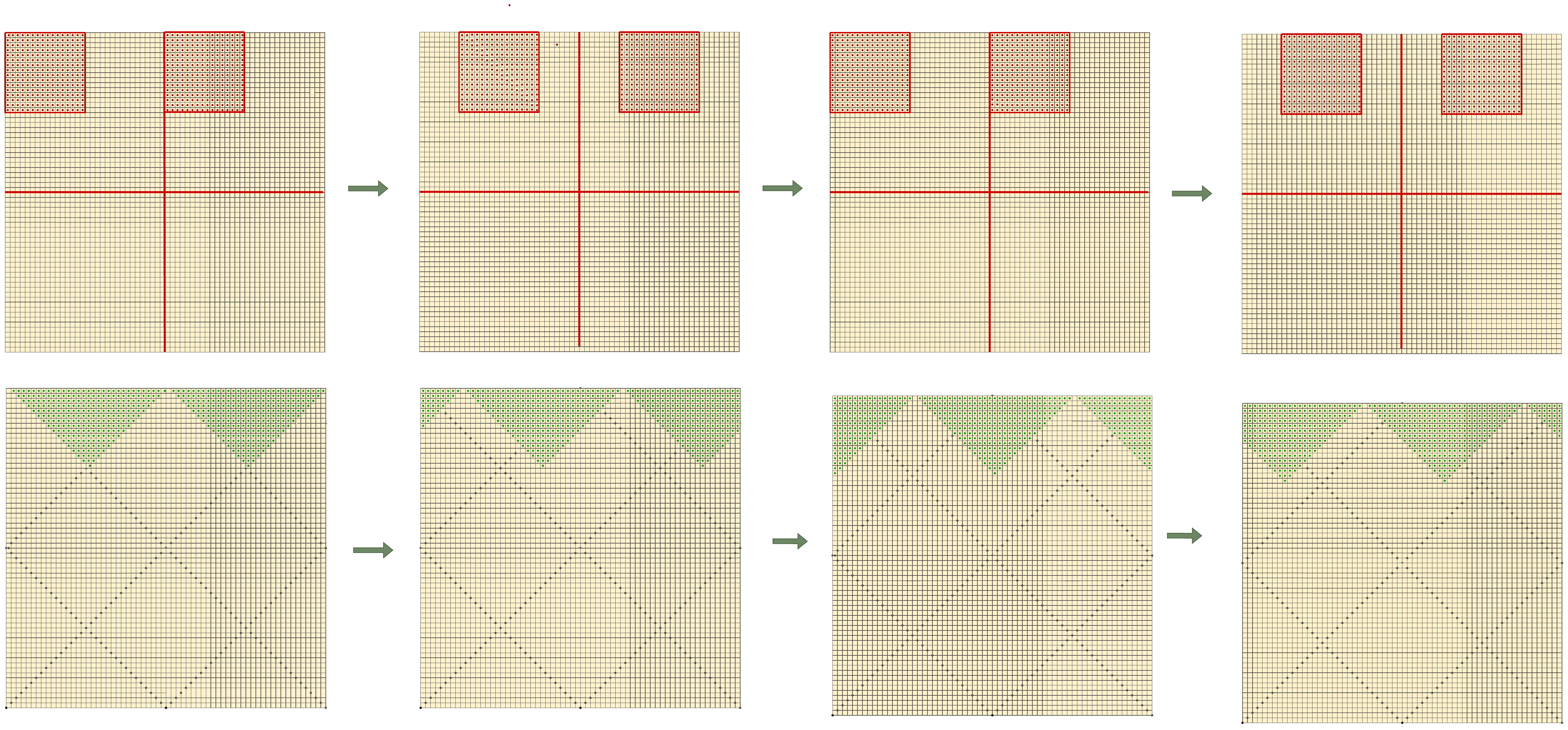}
	\caption{Shifting modes of rectangular and triangular windows in a $64 \times 64$ image patch}
	\label{fig:image2}
	\vspace*{-6mm}
\end{figure}

Many studies confirm that the shifted rectangular window, in conjunction with the non-shifted version, can vastly enhance the SR performance \cite{chen2023activating, conde2022swin2sr, liang2021swinir, zhang2022accurate}. However, we can observe a restriction in shifting the rectangular window as repetition occurs due to its isometric geometry. From Fig.\ref{fig:rect_tri}, it is noticeable that the coverage length of the triangular window is more than the rectangular one in x-dimension. Due to this extended coverage, the triangular windows are allowed more non-identical shifts than rectangular windows. As shown in Fig.\ref{fig:image2}, the rectangular window has non-identical shifting modes of 0 and 8, whereas the triangular window has 0, 8, 16, and 24. The availability of more unique shifting modes in triangular windows enhances the model's performance further than rectangular ones by mitigating the edge-related artifacts at the boundaries, aligning the features that contribute to improving localization accuracy and providing greater adaptability to noncentralized image patterns.

Due to the structural heterogeneity that exists between rectangular and triangular window patches, as shown in Fig. \ref{fig:rect_tri}, the spatial features that participated in the alternative connection of triangular and rectangular self-attention inside DWAB or SWAB are different from each other. These diverse activated features from successive triangular and rectangular attention leverage more diverse spatial features to participate in HR reconstruction. It boosts the performance further.

\vspace*{-4mm}
\paragraph{Computational Cost:} The first step of MSA used in transformer calculates query $q=xw_{q}$, key $k=xw_{k}$ and value $v=xw_{v}$ matrices for an input $x \in R^{HW \times C}$ and the corresponding three trainable parameter matrices $w_{q} \in R^{C \times C_{q}}$, $w_{k} \in R^{C \times C_{k}}$, and $w_{v} \in R^{C \times C_{v}}$. In our case, we consider $C$=$C_{q}$=$C_{k}$=$C_{v}$ and $HW \gg C$. For mathematical convenience, we assume the number of heads equals 1. Considering space complexity for (i) $qk^{T}/\sqrt{C}$ equal to $O([HW]^{2}C)$, (ii) exponential and row sum operations of softmax equal to $O([HW]^2)$, (iii) division operation of softmax equal to $O([HW]^2)$, and value multiplication operation equal to $O([HW]^{2}C$, the approximate computational complexity of MSA is
\vspace*{-2mm}
\begin{equation}
	O(MSA) = 4[HW]C^{2} + 2[HW]^{2}C
	\vspace*{-2mm}
\end{equation}
The above equation confirms the high computational cost of deploying ViTs. The rectangular window technique is the only solution to lower the complexity so far. However, we proposed a triangular window technique that not only reduces the computational burden but also yields better results than a rectangular counterpart while working alone and even produces superior performance while employed synchronously with a rectangular window. In this paper, we introduce two variants of attention relating to the proposed triangular window: (i) dense triangular attention and (ii) sparse triangular attention.

The dense triangular attention has a 1D token size of $L^2$, where $L$ is the triangular window equivalent of rectangular window token $L \times L$. The computational cost of the proposed dense TW-MSA is
\vspace*{-3mm}
\begin{equation}
	\label{eq:DTW-MSA}
	O(\text{D-MSA}) = (4HWC^{2} + 2HWL^{2}C)
	\vspace*{-3mm}
\end{equation}

The effective sequence length in sparse triangular attention is the same as dense but for a wider receptive field. The interval length $S$ controls the sparsity in the triangular window during the attention mechanism. $S$ can be compared with the stride in convolution operation. The computational cost of sparse TW-MSA in terms of parameter $S$ is
\vspace*{-3mm}
\begin{equation}
	\label{eq:STW-MSA}
	O(\text{S-MSA}) = (4HWC^{2} + 2(\frac{HW}{S})^{2}C)
	\vspace*{-3mm}
\end{equation}
The computational cost of dense TW-MSA and sparse TW-MSA are drastically improved as $L^{2} \ll HW$ in equation \ref{eq:DTW-MSA} and $(\frac{HW}{S})^{2} \ll (HW)^{2}$ in equation \ref{eq:STW-MSA}, respectively.

\vspace*{-2mm}
\section{Experiments}
\label{sec:Experiments}

\subsection{Datasets and Performance Matrices}
\vspace*{-1mm}
We train the proposed model with two widely used DIV2K\cite{timofte2018challenge} and Flicker2K \cite{lim2017enhanced} datasets to avoid overfitting while training only on DIV2K and evaluate on standard SR benchmarks: Set5\cite{bevilacqua2012low}, Set14\cite{zeyde2012single}, BSD100\cite{martin2001database}, Urban100\cite{huang2015single}, and Manga109\cite{matsui2017sketch}. We evaluate the proposed SR model using PSNR and SSIM on the Y channel in YCbCr Space.  
\vspace*{-2mm}
\subsection{Experimental Settings}
\vspace*{-1mm}
We perform geometric data augmentation where each training sample is replaced by its flipped or rotated ($90^\circ$, $180^\circ$, and $270^\circ$) equivalent. The augmented input samples are then cropped into 64$\times$64 patches before passing through the model. While configuring the proposed CFAT, we put 3 DWAB and 3 SWAB in an alternative order. We maintain 4 D-HWAB and 4 S-HWAB inside each DWAB and SWAB unit, respectively. Here, each D-HWAB and S-HWAB unit represents a unique shift size, i.e., 0, 8, 16, and 24. We take 180 channels, while attention heads and window sizes are set to 6 and 16 for all units of (SD)RW-MSA, (S)TW-MSA, and OCFA. Other hyperparameter values are discussed in the ablation study. We set the batch size to 32. We use ADAM optimizer for model training with $\beta_1$ = 0.9 and $\beta_2$ = 0.999. The weight decay is set to zero. The total iteration during training is set to 250K. The initial learning rate is set to 2e-4, which is reduced to half after [112.5K, 175K, 200K, 225K] iterations. We use $L_1$ loss for model training. We take the pre-trained $ \times 2$ model and finetune it for $\times 4$ resolution. We also introduced a more compact CFAT, namely CFAT-S, with 144 channels, featuring a depthwise-pointwise convolution in CWAB. The experiments are performed using Python 3.10.11 based PyTorch 2.0.1 deep learning framework on an Ubuntu 20.04.2 machine with A100-PCIE-40GB GPU enabled with Nvidia CUDA 10.1.243 and CuDNN 8.1.0. 
\vspace*{-1mm}
\begin{table}[h]
	\vspace*{-2mm}
	\begin{center}
		\tabcolsep=0.13cm
		\centering
		\captionsetup{justification=centering}
		\caption{Performance based on various triangular \& rectangular window size, and shift size. L: Linear, S: Square.}
		\vspace*{-6.5mm}
		\singlespacing
		\scalebox{0.7}{
			\begin{tabular}{cccccc}
				\toprule
				\toprule
				Triangular window size & 8*8 (L) & 16*16 (L) &&&\\
				\midrule
				PSNR (in dB)  & 27.58 & 27.77 & && \rule{0pt}{1ex}\\
				\midrule
				\midrule  
				Rectangular window Size  & 8 (S) & 12 (S) & 16 (S)&& \rule{0pt}{1ex} \\
				\midrule 
				PSNR (in dB)  & 27.52 & 27.64 & 27.77 && \\ 
				\midrule
				\midrule  
				Shift Size &0  & 8 & 16 & 24 & all\rule{0pt}{1ex} \\
				\midrule 
				PSNR (in dB) & 27.58 & 27.68 & 27.67 & 27.66 & 27.77 \\   
				\bottomrule
				\bottomrule
		\end{tabular}}
		\label{table:shift_size}
		\vspace*{-5mm}
	\end{center}
\end{table}
\begin{table}[h]
	\vspace*{-2mm}
	\begin{center}
		\tabcolsep=0.13cm
		\vspace*{-2mm}
		\captionsetup{justification=centering}
		\caption{Ablation study for $\alpha$, $\beta$, and channel counts.}
		\vspace*{-6.5mm}
		\singlespacing
		\scalebox{0.7}{
			\begin{tabular}{ccccc}
				\toprule
				\toprule  
				\(\alpha\)  & 0 & 1 & 0.1 & 0.01 \rule{0pt}{1ex} \\
				\midrule 
				PSNR (in dB)  & 27.70 & 27.67 & 27.61 & 27.77 \\   
				\midrule
				\midrule  
				\(\beta\)  & 0 & 1 & 0.1 & 0.015 \rule{0pt}{1ex} \\
				\midrule 
				PSNR (in dB)   & 27.72 & 27.65 & 27.69 & 27.78  \\ 
				\midrule 
				\midrule 
				Channel Count  & 180 & 144 & 96 \rule{0pt}{1ex}& \\
				\midrule
				PSNR (in dB)  & 27.77 & 27.45 & 27.21 & \\
				\bottomrule
				\bottomrule
		\end{tabular}}
		\label{table:abchannel}
		
	\end{center}
	\vspace*{-5mm}
\end{table}

\begin{table}[h]
	\vspace*{-2mm}
	\begin{center}
		\tabcolsep=0.13cm
		\vspace*{-2mm}
		\captionsetup{justification=centering}
		\caption{Performance based on overlapping constant.}
		\vspace*{-6.5mm}
		\singlespacing
		\scalebox{0.7}{
			\begin{tabular}{ccccc}
				
				\toprule
				\toprule
				$I$ & 0 & 2 & 4 & 6 \\
				\midrule
				PSNR(dB)  & 27.75  & 27.76   & 28.72    & 27.70 \rule{0pt}{1ex}   \\
				\midrule
				\midrule
				\(k\) & 0 & 0.25 & 0.5 & 0.75 \\
				\midrule
				PSNR(dB)  & 27.68  & 27.65   & 27.77    & 27.72    \\
				\bottomrule
				\bottomrule
		\end{tabular}}
		\label{table:IKvalue}
	\end{center}
	\vspace*{-5mm}
\end{table}

\begin{table}[h]
	\vspace*{-2mm}
	\begin{center}
		\tabcolsep=0.13cm
		\vspace*{-2mm}
		\captionsetup{justification=centering}
		\caption{Ablation study for CWAB, RWAB, and TWAB.}
		\vspace*{-6.5mm}
		\singlespacing
		\scalebox{0.7}{
			\begin{tabular}{cccc}
				\toprule
				\toprule
				Structure & w/o CWAB & w/ CWAB & CWAB with DP conv \rule{0pt}{1ex} \\
				\midrule
				PSNR/SSIM  & 27.65dB/0.7481 & 27.75dB/0.7489 & 27.77dB/0.7494 \\
				\midrule
				\midrule
				Structure & w/o RWAB & w/o TWAB & w/ RWAB \& TWAB  \rule{0pt}{4ex}\\
				\midrule
				PSNR/SSIM  & 27.61dB/0.7463 & 27.56dB/0.7471 & 27.77dB/0.7494\\ 
				\bottomrule
				\bottomrule
		\end{tabular}}
		\label{table:CWTWRW}
	\end{center}
	\vspace*{-9mm}
\end{table}
\subsection{Ablation Study}
\vspace*{-1.5mm}
\subsubsection{Impact of Various Model Hyperparameters :}
\vspace*{-2mm}
To realize the impact of various model hyperparameters, we conduct all our experiments on BSD100 dataset for a scale of $\times 4$ and a epoch of 70 and compute the model's performance in terms of PSNR.
\vspace*{-5mm}
\paragraph{Window Size and Shift Size :}
In \cite{chen2023activating}, the authors reveal that optimum rectangular window size can activate more pixels that in turn elevate model performances. \cite{li2021efficient} and \cite{chen2023activating} investigate their model with square patch sizes of (8, 12) and (8, 16) respectively. In this paper, we inspect our model on the square patch of (8, 12, 16) for the rectangular and linear patch of (8*8, 16*16) for triangular windows. The experimental results are shown in Tab. \ref{table:shift_size}. Again, we evaluate the performance with (8, 16, 24, variable) shift and without shift (0), which is also mentioned in this table. Here, the 'variable' refers to varying shift size from $0\rightarrow 8\rightarrow16\rightarrow24$ in succession. From the above, we can conclude that the model yields the best PSNR for $16*16$ triangular linear-window, $16 \times 16$ rectangular square-window, and variable shift size. 
\vspace*{-5mm}
\paragraph{Interval Size :} Interval size ($I$) plays a decisive role in estimating model performance \cite{zhang2022accurate}. A smaller $I$ refers to the wide receptive field with high computation, and a larger $I$ implies the opposite. However, we prioritize the performance and compute model results by varying the $I$. We can observe from Tab. \ref{table:IKvalue} that the model gives the best performances for interval sizes 0 and 2. However, we chose $I=2$ as it allows us to design our model with fewer computations.
\vspace*{-5mm}

\paragraph{$\alpha, \beta$, Channel Counts and CWAB:}
To examine the influence of weighing factor $\alpha$ and $\beta$, we measure the model outcome for four different values (0, 1, 0.1, 0.01) of $\alpha$ and $\beta$ separately. As displayed in Tab. \ref{table:abchannel}, these experiments specify the optimum value of $\alpha$ and $\beta$ as 0.01 and 0.015, respectively. We also scrutinize the effect of channel count in CFAT as shown in Tab. \ref{table:abchannel}. We select 180 as the channel counts for CFAT as the rise in channel non-linearly increases the model parameters, and also, the model performance slowly gets saturated after 180.
\vspace*{-5mm}
\paragraph{Overlapping Constant :}
In OCFAB, we use a constant $k$ to determine the overlapping range between two consecutive windows during cross-attention. We study the impact of various overlapping ratios by adjusting $k$ values ranging from 0 to 0.75, and the results are presented in Tab. \ref{table:IKvalue}. Here, $k=0$ corresponds to a standard Transformer block. The results displayed in this table indicate that the model delivers optimal performance for $k=0.5$.
\vspace*{-5mm}
\subsubsection{Impact of CWAB :}
\vspace*{-1mm}
We have performed three experiments to study the importance of CWAB. From Tab. \ref{table:CWTWRW}, it is quite evident that the PSNR value of CFAT without CWAB is 0.1dB lower than CFAT using CWAB with standard convolution. However, we get a slight performance advantage when it comes to model outcomes with CWAB using depthwise-pointwise (DP) convolutions and lower squeeze factor. So, we get a better result without a computational trade-off, as the latter two have the same complexity.
\vspace*{-5mm}
\subsubsection{Impact of Triangular and Rectangular Windows:}
\vspace*{-1mm}
In CFAT, RWAB and TWAB modules carry significant roles in CFAT realization. From Tab. \ref{table:CWTWRW}, we can observe a significant drop of $0.21\sim0.16$ in PSNR for our model without TWAB or RWAB. These results confirm our design of including both units in the final model. They combinely truncate the effect of edge-level distortion that leads to enhanced performance. In this table, we also observe a narrow advantage for TWAB over RWAB when individually evaluated due to the feasibility of extra shifting modes.
\vspace*{-2.5mm}
\begin{table*}[t]
	\vspace*{-3mm}
	\captionsetup{justification=centering}
	\caption{Quantitative comparison of the CFAT with various state-of-the-art SR methods.\\ \textcolor{red}{red} and \textcolor{green}{green} color indicate the  best and second best respectively}
	\vspace*{-9.5mm}
	\begin{center}
		\tabcolsep=0.11cm
		\singlespacing
		\scalebox{0.7}{
			\begin{tabular}{lcccccccccccc}
				\toprule
				\multirow{2}{*}{\textbf{Method}} &
				\multirow{2}{*}{\textbf{Scale}} &
				\multirow{2}{*}{\textbf{Training Dataset}} &
				\multicolumn{2}{c}{\textbf{Set5}} &
				\multicolumn{2}{c}{\textbf{Set14}} &
				\multicolumn{2}{c}{\textbf{BSD100}} &
				\multicolumn{2}{c}{\textbf{Urban100}} &
				\multicolumn{2}{c}{\textbf{Manga109}}  \\
				\cmidrule(lr){4-5}\cmidrule(lr){6-7}\cmidrule(lr){8-9}\cmidrule(lr){10-11}\cmidrule(lr){12-13}
				&&& {PSNR} & {SSIM} & {PSNR} & {SSIM} & {PSNR} & {SSIM} & {PSNR} & {SSIM} & {PSNR} & {SSIM} \\
				\midrule
				\midrule
				EDSR \cite{lim2017enhanced} & $\times 2$ & DIV2K & 38.11 & 0.9602 & 33.92 & 0.9195 & 32.32 & 0.9013 & 32.93 & 0.9351 & 39.10 & 0.9773\\
				HAN \cite{niu2020single} & $\times 2$ & DIV2K & 38.27 & 0.9614 & 34.16 & 0.9217 & 32.41 & 0.9027 & 33.35 & 0.9385 & 39.46 & 0.9785\\
				SAN \cite{dai2019second} & $\times 2$ & DIV2K & 38.31 & 0.9620 & 34.07 & 0.9213 & 32.42 & 0.9028 & 33.10 & 0.9370 & 39.32 & 0.9792\\
				\hdashline
				IPT \cite{chen2021pre} & $\times 2$ & ImageNet & 38.37 & - & 34.43 & - & 32.48 & - & 33.76 & - & - & -\\
				SwinIR  \cite{liang2021swinir} & $\times 2$ & DIV2K+Flickr2K & 38.46  & 0.9624 & 34.61 & 0.9260 & 32.55 & 0.9043 & 33.95 & 0.9433 & 40.02 & 0.9800\\
				Swin2SR \cite{conde2022swin2sr} & $\times 2$ & DIV2K+Flickr2K & 38.43 & 0.9623 & 34.48 & 0.9256 & 32.54 & 0.905 & 33.89 & 0.9431 & 39.88 & 0.9798  \\
				ACT \cite{yoo2023enriched} & $\times 2$ & DIV2K+Flickr2K & 38.53 & 0.9629 & 34.68 & 0.9260 & 32.60 & 0.9052 & 34.25 & 0.9453 & 40.11 & 0.9807\\
				ART \cite{zhang2022accurate} & $\times 2$ & DIV2K+Flickr2K & 38.56 & 0.9629 & 34.59 & 0.9267 & 32.58 & 0.9048 & 34.30 & 0.9452 & 40.24 & 0.9808\\
				EDT  \cite{li2021efficient} & $\times 2$ & DIV2K+Flickr2K & 38.63 & \textcolor{green}{0.9632} & 34.80 & 0.9273 & 32.62 & 0.9052 & 34.27 & 0.9456 & \textcolor{green}{40.37} & \textcolor{green}{0.9811} \\
				HAT \cite{chen2022activating} & $\times 2$ & DIV2K+Flickr2K & \textcolor{green}{38.63} & 0.9630 & \textcolor{green}{34.86} & \textcolor{green}{0.9274} & \textcolor{green}{32.62} & \textcolor{green}{0.9053} & \textcolor{green}{34.45} & \textcolor{green}{0.9466} & 40.26 & 0.9809\\
				CFAT-s (ours) & $\times 2$ & DIV2K+Flickr2K & 38.59 & 0.9621 & 34.81 & 0.92872 & 32.58 & 0.9044 & 34.42 & 0.9453 & 40.24 & 0.9799 \rule{0pt}{3ex} \\
				CFAT (ours) & $\times 2$ & DIV2K+Flickr2K & \textcolor{red}{39.09} & \textcolor{red}{0.9653} & \textcolor{red}{35.25} & \textcolor{red}{0.9296} & \textcolor{red}{32.93} & \textcolor{red}{0.9072} & \textcolor{red}{35.01} & \textcolor{red}{0.9498} & \textcolor{red}{41.00} & \textcolor{red}{0.9838}  \\
				\hline
				
				EDSR \cite{lim2017enhanced} & $\times 3$ & DIV2K & 34.65 & 0.9280 & 30.52 & 0.8462 & 29.25 & 0.8093 & 28.80 & 0.8653 & 34.17 & 0.9476 \\
				HAN \cite{niu2020single} & $\times 3$ & DIV2K & 34.75 &  0.9299 & 30.67 & 0.8483 & 29.32 & 0.8110 & 29.10 & 0.8705 & 34.48 & 0.9500\\
				SAN \cite{dai2019second} & $\times 3$ & DIV2K & 34.75 & 0.9300 & 30.59 & 0.8476 & 29.33 & 0.8112 & 28.93 & 0.8671 & 34.30 & 0.9494\\
				\hdashline
				IPT \cite{chen2021pre} & $\times 3$ & ImageNet & 34.81 & - & 30.85 & - & 29.38 & - & 29.49 & - & - & -\\
				SwinIR  \cite{liang2021swinir} & $\times 3$ & DIV2K+Flickr2K & 35.04 & 0.9322 & 31.00 & 0.8542 & 29.49 & 0.8150 & 29.90 & 0.8841 & 35.28 & 0.9543 \\
				ACT \cite{yoo2023enriched} & $\times 3$ & DIV2K+Flickr2K & 35.09 & 0.9325 & 31.17 & 0.8549 & 29.55 & 0.8171 & 30.26 & 0.8876 & 35.47 & 0.9548\\
				ART \cite{zhang2022accurate} & $\times 3$ & DIV2K+Flickr2K & 35.07 & 0.9325 & 31.02 & 0.8541 & 29.51 & 0.8159 & 30.10 & 0.8871 & 35.39 & 0.9548\\
				EDT  \cite{li2021efficient} & $\times 3$ & DIV2K+Flickr2K & \textcolor{green}{35.13} &  0.9328 & \textcolor{green}{31.09} & 0.8553 & 29.53 & 0.8165 & 30.07 & 0.8863 & 35.47 & 0.9550 \\
				HAT \cite{chen2022activating} & $\times 3$ & DIV2K+Flickr2K & 35.07 & \textcolor{green}{0.9329} & 31.08 & \textcolor{green}{0.8555} & \textcolor{green}{29.54} & \textcolor{green}{0.8167} & \textcolor{green}{30.23} & \textcolor{green}{0.8896} & \textcolor{green}{35.53} & \textcolor{green}{0.9552}  \\
				CFAT-s (ours) & $\times 3$ & DIV2K+Flickr2K & 34.03 & 0.9323 & 31.06 & 0.8551 & 29.57 & 0.8158 & 30.18 & 0.8889 & 35.48 & 0.9547 \rule{0pt}{3ex} \\
				CFAT (ours) & $\times 3$ & DIV2K+Flickr2K & \textcolor{red}{35.31} & \textcolor{red}{0.9340} & \textcolor{red}{31.32} & \textcolor{red}{0.8569} & \textcolor{red}{29.70} & \textcolor{red}{0.8180} & \textcolor{red}{30.43} & \textcolor{red}{0.8928} & \textcolor{red}{35.82} & \textcolor{red}{0.9574}  \\
				\hline
				
				EDSR \cite{lim2017enhanced} & $\times 4$ & DIV2K & 32.46 & 0.8968 & 28.80 & 0.7876 & 27.71 & 0.7420 & 26.64 & 0.8033 & 31.02 & 0.9148\rule{0pt}{3ex}\\
				HAN \cite{niu2020single} & $\times 4$ & DIV2K & 32.64 & 0.9002 & 28.90 & 0.7890 & 27.80 & 0.7442 & 26.85 & 0.8094 & 31.42 & 0.9177\\
				SAN \cite{dai2019second} & $\times 4$ & DIV2K & 32.64 & 0.9003 & 28.92 & 0.7888 & 27.78 & 0.7436 & 26.79 & 0.8068 & 31.18 & 0.9169\\
				\hdashline
				IPT \cite{chen2021pre} & $\times 4$ & ImageNet & 32.64 & - & 29.01 & - & 27.82 & - & 27.26 & - & - & - \\
				SwinIR  \cite{liang2021swinir} & $\times 4$ & DIV2K+Flickr2K & 32.93 & 0.9043 & 29.15 & 0.7958 & 27.95 & 0.7494 & 27.56 & 0.8273 & 32.22 & 0.9273\\
				Swin2SR \cite{conde2022swin2sr} & $\times 4$ & DIV2K+Flickr2K & 32.92 & 0.9039 & 29.06 & 0.7946 & 27.92 & 0.7505 & 27.51 & 0.8271 & 31.03 & 0.9256 \\
				ACT \cite{yoo2023enriched} & $\times 4$ & DIV2K+Flickr2K & 33.04 & 0.9041 & 29.27 & 0.7968 & 28.00 & 0.7516 & 27.92 & 0.8332 & 32.44 & 0.9282\\
				ART \cite{zhang2022accurate} & $\times 4$ & DIV2K+Flickr2K & 33.04 & 0.9051 & 29.16 & 0.7958 & 27.97 & 0.7510 & 27.77 & 0.8321 & 32.31 & 0.9283\\
				EDT  \cite{li2021efficient} & $\times 4$ & DIV2K+Flickr2K & \textcolor{green}{33.06} & 0.9055 & 29.23 & 0.7971 & 27.99 & 0.7510 & 27.75 & 0.8317 & 32.39 & 0.9283 \\
				HAT \cite{chen2022activating} & $\times 4$ & DIV2K+Flickr2K & 33.04 & \textcolor{green}{0.9056} & \textcolor{green}{29.23} & \textcolor{green}{0.7973} & \textcolor{green}{28.00} & \textcolor{green}{0.7517} & \textcolor{green}{27.97} & \textcolor{green}{0.8368} & \textcolor{green}{32.48} & \textcolor{green}{0.9292}  \\
				CFAT-s (ours) & $\times 4$ & DIV2K+Flickr2K & 32.01 & 0.9045 & 29.25 & 0.7972 & 27.99 & 0.7504 & 27.86 & 0.8358 & 32.45 & 0.9279  \rule{0pt}{3ex} \\
				CFAT (ours) & $\times 4$ & DIV2K+Flickr2K & \textcolor{red}{33.19} & \textcolor{red}{0.9068} & \textcolor{red}{29.30} & \textcolor{red}{0.7985} & \textcolor{red}{28.17} & \textcolor{red}{0.7524} & \textcolor{red}{28.11} & \textcolor{red}{0.8380} & \textcolor{red}{32.63} & \textcolor{red}{0.9305}  \\
				\bottomrule
				\bottomrule
		\end{tabular}}
		\label{table:table7}
	\end{center}
	\vspace*{-7mm}
\end{table*}

\subsection{Comparisons with State-of-the-art Methods }
\vspace*{-1mm}
\paragraph{Quantitative Analysis :} We select CNN-based EDSR\cite{lim2017enhanced}, SAN\cite{dai2019second}, and HAN\cite{niu2020single} with transformer-based IPT\cite{chen2021pre}, SwinIR\cite{liang2021swinir}, Swin2SR\cite{conde2022swin2sr}, ACT\cite{yoo2023enriched}, ART\cite{zhang2022accurate}, EDT\cite{li2021efficient}, HAT\cite{chen2023activating} state-of-the-art architectures to compare our model quantitatively in term of PSNR and SSIM. Tab. \ref{table:table7}  presents the quantitative comparison of CFAT based on scale factors $\times 2$, $\times 3$, and $\times 4$ for both DIV2K and DF2K (DIV2K+Flickr2K) training datasets. HAT\cite{chen2023activating} is considered the previous best model whose outcome outplays any other SOTA models. However, due to the undistort and rich feature exploration capability of CFAT, it yields superior performances than HAT \cite{chen2023activating} for all scale factors across all five benchmark datasets. The superior results confirm the significance of using triangular window attention in CFAT. The small variant of CFAT, called CFAT-s, is also very competitive while comparing its outcomes with some well-known transformer-based architectures like IPT \cite{chen2021pre}, SwinIR \cite{liang2021swinir} and Swin2SR \cite{conde2022swin2sr}.

\begin{figure*}[t]
	\centering
	\captionsetup{justification=centering}
	\includegraphics[width=0.995\linewidth]{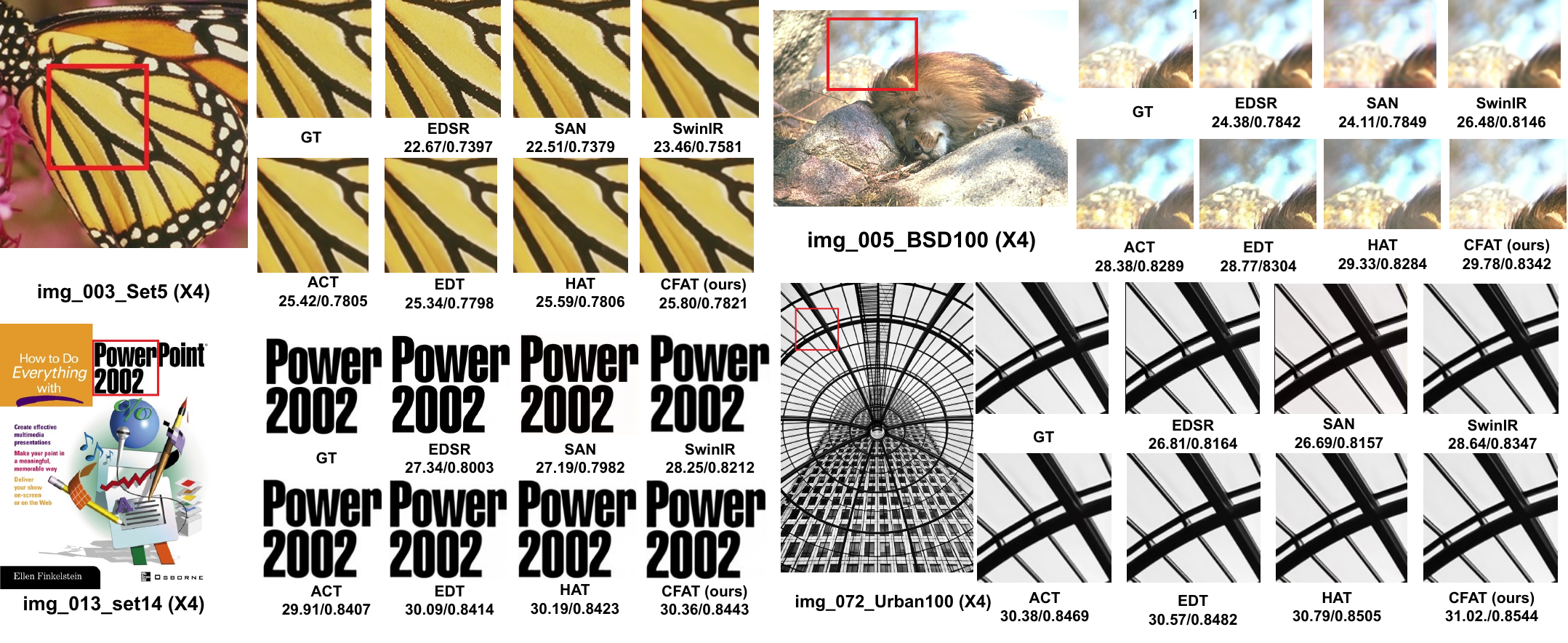}
	\vspace*{-5mm}
	\caption{Visual Comparison of CFAT with other state-of-the-art methods.}
	\label{fig:image3}
	\vspace*{-7mm}
\end{figure*}
\vspace*{-4.5mm}
\paragraph{Visual Analysis :}
For visual assessment, we select an image from each of the four distinct testing datasets: ``img\_003" from Set5, ``img\_072" from Urban100, ``img\_013" from Set14, and ``img\_005" from BSD100. CFAT retrieves all these images with less blur and fair lattice content, as shown in Fig.\ref{fig:image3}. The PSNR and SSIM value for each image in this figure confirms our model's superiority over all SOTA models. Our model reduces the pixel level distortion more efficiently than any other CNN- and transformer-based leading models due to the inclusion of dense and sparse triangular window attention.
\vspace*{-4.5mm}
\paragraph{Comparison based on Computational Cost :} 
From Tab. \ref{table:comp_cost}, we observe that the performance of CFAT outperforms other SOTA models with a balanced trade-off between parameter and multi-add count. Our model has a few more parameters than HAT \cite{chen2023activating} but shows fewer counts of multi-adds while giving superior results on the BSD100 dataset.
\vspace*{-7mm}
\begin{table}[H]
	\begin{center}
		\caption{Model comparison based on computational cost}
		\vspace*{-7mm}
		\tabcolsep=0.13cm
		\singlespacing
		\scalebox{0.8}{
			\begin{tabular}{cccc}
				\toprule 
				\toprule 
				Methods & Params (M) & Multi-adds(GMac) & PSNR/SSIM \rule{0pt}{1ex} \\
				\midrule
				SwinIR  & 11.9 & 53.6 & 27.92dB/0.7489 \\
				\midrule
				ACT  & 46 & 22 & 28.00dB/0.7516 \\
				\midrule
				ART  & 16.55  & 120 & 27.97dB/0.7510 \\
				\midrule
				EDT  & 11.6 & - & 27.91dB/0.7483 \\
				\midrule
				HAT  & 20.8  & 103.7 & 28.00dB/0.7517 \\
				\midrule
				CFAT  & 22.07  & 90.59 & 28.17dB/0.7524 \\
				\bottomrule
				\bottomrule
		\end{tabular}}
		\label{table:comp_cost}
	\end{center}
	\vspace*{-7mm}
\end{table}

\section{Conclusion}
\vspace*{-2mm}
We propose a rectangular and triangular window attention-based SR architecture called Composite Fusion Attention Transformer (CFAT). This combinational design eliminates boundary-level distortion problems and opens the gate for integrating more non-identical shifting modes. Dense and sparse attention in both windows allows interaction between local and global image features on the same platform. We also combine more diverse features through non-overlapping triangular and rectangular window-based self-attention and overlapping window-based cross-attention. Extensive experiments on multiple datasets show the effectiveness of CFAT. By incorporating the novel triangular window attention in dense, sparse, and shifted configuration, CFAT outperforms the other state-of-the-art models qualitatively and quantitatively.

{\small
\bibliographystyle{ieee_fullname}
\bibliography{egbib}

\begin{thebibliography}{10}\itemsep=-1pt

\bibitem{bevilacqua2012low}
Marco Bevilacqua, Aline Roumy, Christine Guillemot, and Marie~Line
  Alberi-Morel.
\newblock Low-complexity single-image super-resolution based on nonnegative
  neighbor embedding.
\newblock 2012.

\bibitem{cao2022swin}
Hu Cao, Yueyue Wang, Joy Chen, Dongsheng Jiang, Xiaopeng Zhang, Qi Tian, and
  Manning Wang.
\newblock Swin-unet: Unet-like pure transformer for medical image segmentation.
\newblock In {\em European conference on computer vision (ECCV)}, pages
  205--218. Springer, 2022.

\bibitem{carion2020end}
Nicolas Carion, Francisco Massa, Gabriel Synnaeve, Nicolas Usunier, Alexander
  Kirillov, and Sergey Zagoruyko.
\newblock End-to-end object detection with transformers.
\newblock In {\em European conference on computer vision (ECCV)}, pages
  213--229. Springer, 2020.

\bibitem{chen2021pre}
Hanting Chen, Yunhe Wang, Tianyu Guo, Chang Xu, Yiping Deng, Zhenhua Liu, Siwei
  Ma, Chunjing Xu, Chao Xu, and Wen Gao.
\newblock Pre-trained image processing transformer.
\newblock In {\em Computer Vision and Pattern Recognition (CVPR)}, pages
  12299--12310. IEEE/CVF, 2021.

\bibitem{chen2022activating}
Xiangyu Chen, Xintao Wang, Jiantao Zhou, and Chao Dong.
\newblock Activating more pixels in image super-resolution transformer. arxiv
  2022.
\newblock {\em arXiv preprint arXiv:2205.04437}, 1, 2022.

\bibitem{chen2023activating}
Xiangyu Chen, Xintao Wang, Jiantao Zhou, Yu Qiao, and Chao Dong.
\newblock Activating more pixels in image super-resolution transformer.
\newblock In {\em Computer Vision and Pattern Recognition (CVPR)}, pages
  22367--22377. IEEE/CVF, 2023.

\bibitem{chu2021twins}
Xiangxiang Chu, Zhi Tian, Yuqing Wang, Bo Zhang, Haibing Ren, Xiaolin Wei,
  Huaxia Xia, and Chunhua Shen.
\newblock Twins: Revisiting the design of spatial attention in vision
  transformers.
\newblock {\em Advances in Neural Information Processing Systems (NeurIPS)},
  34:9355--9366, 2021.

\bibitem{conde2022swin2sr}
Marcos~V Conde, Ui-Jin Choi, Maxime Burchi, and Radu Timofte.
\newblock Swin2sr: Swinv2 transformer for compressed image super-resolution and
  restoration.
\newblock In {\em European conference on computer vision (ECCV)}, pages
  669--687. Springer, 2022.

\bibitem{dai2019second}
Tao Dai, Jianrui Cai, Yongbing Zhang, Shu-Tao Xia, and Lei Zhang.
\newblock Second-order attention network for single image super-resolution.
\newblock In {\em Computer Vision and Pattern Recognition (CVPR)}, pages
  11065--11074. IEEE/CVF, 2019.

\bibitem{dong2014learning}
Chao Dong, Chen~Change Loy, Kaiming He, and Xiaoou Tang.
\newblock Learning a deep convolutional network for image super-resolution.
\newblock In {\em European conference on computer vision (ECCV)}, pages
  184--199. Springer, 2014.

\bibitem{dong2015image}
Chao Dong, Chen~Change Loy, Kaiming He, and Xiaoou Tang.
\newblock Image super-resolution using deep convolutional networks.
\newblock {\em IEEE transactions on pattern analysis and machine intelligence
  (PAMI)}, 38(2):295--307, 2015.

\bibitem{dong2016accelerating}
Chao Dong, Chen~Change Loy, and Xiaoou Tang.
\newblock Accelerating the super-resolution convolutional neural network.
\newblock In {\em European conference on computer vision (ECCV)}, pages
  391--407. Springer, 2016.

\bibitem{dosovitskiy2020image}
Alexey Dosovitskiy, Lucas Beyer, Alexander Kolesnikov, Dirk Weissenborn,
  Xiaohua Zhai, Thomas Unterthiner, Mostafa Dehghani, Matthias Minderer, Georg
  Heigold, Sylvain Gelly, et~al.
\newblock An image is worth 16x16 words: Transformers for image recognition at
  scale.
\newblock {\em arXiv preprint arXiv:2010.11929}, 2020.

\bibitem{gu2021interpreting}
Jinjin Gu and Chao Dong.
\newblock Interpreting super-resolution networks with local attribution maps.
\newblock In {\em Computer Vision and Pattern Recognition (CVPR)}, pages
  9199--9208. IEEE/CVF, 2021.

\bibitem{guo2020closed}
Yong Guo, Jian Chen, Jingdong Wang, Qi Chen, Jiezhang Cao, Zeshuai Deng, Yanwu
  Xu, and Mingkui Tan.
\newblock Closed-loop matters: Dual regression networks for single image
  super-resolution.
\newblock In {\em Computer Vision and Pattern Recognition (CVPR)}, pages
  5407--5416. IEEE/CVF, 2020.

\bibitem{huang2022glance}
Gao Huang, Yulin Wang, Kangchen Lv, Haojun Jiang, Wenhui Huang, Pengfei Qi, and
  Shiji Song.
\newblock Glance and focus networks for dynamic visual recognition.
\newblock {\em IEEE Transactions on Pattern Analysis and Machine Intelligence
  (PAMI)}, 45(4):4605--4621, 2022.

\bibitem{huang2015single}
Jia-Bin Huang, Abhishek Singh, and Narendra Ahuja.
\newblock Single image super-resolution from transformed self-exemplars.
\newblock In {\em Computer Vision and Pattern Recognition (CVPR)}, pages
  5197--5206. IEEE/CVF, 2015.

\bibitem{hui2018fast}
Zheng Hui, Xiumei Wang, and Xinbo Gao.
\newblock Fast and accurate single image super-resolution via information
  distillation network.
\newblock In {\em Computer Vision and Pattern Recognition (CVPR)}, pages
  723--731. IEEE/CVF, 2018.

\bibitem{kim2016accurate}
Jiwon Kim, Jung~Kwon Lee, and Kyoung~Mu Lee.
\newblock Accurate image super-resolution using very deep convolutional
  networks.
\newblock In {\em Computer Vision and Pattern Recognition (CVPR)}, pages
  1646--1654. IEEE/CVF, 2016.

\bibitem{kong2021classsr}
Xiangtao Kong, Hengyuan Zhao, Yu Qiao, and Chao Dong.
\newblock Classsr: A general framework to accelerate super-resolution networks
  by data characteristic.
\newblock In {\em Computer Vision and Pattern Recognition (CVPR)}, pages
  12016--12025. IEEE/CVF, 2021.

\bibitem{ledig2017photo}
Christian Ledig, Lucas Theis, Ferenc Husz{\'a}r, Jose Caballero, Andrew
  Cunningham, Alejandro Acosta, Andrew Aitken, Alykhan Tejani, Johannes Totz,
  Zehan Wang, et~al.
\newblock Photo-realistic single image super-resolution using a generative
  adversarial network.
\newblock In {\em Computer Vision and Pattern Recognition (CVPR)}, pages
  4681--4690. IEEE/CVF, 2017.

\bibitem{li2022hst}
Bingchen Li, Xin Li, Yiting Lu, Sen Liu, Ruoyu Feng, and Zhibo Chen.
\newblock Hst: Hierarchical swin transformer for compressed image
  super-resolution.
\newblock In {\em European Conference on Computer Vision (ECCV)}, pages
  651--668. Springer, 2022.

\bibitem{li2023uniformer}
Kunchang Li, Yali Wang, Junhao Zhang, Peng Gao, Guanglu Song, Yu Liu, Hongsheng
  Li, and Yu Qiao.
\newblock Uniformer: Unifying convolution and self-attention for visual
  recognition.
\newblock {\em IEEE Transactions on Pattern Analysis and Machine Intelligence
  (PAMI)}, 2023.

\bibitem{li2021efficient}
Wenbo Li, Xin Lu, Jiangbo Lu, Xiangyu Zhang, and Jiaya Jia.
\newblock On efficient transformer and image pre-training for low-level vision.
\newblock {\em arXiv preprint arXiv:2112.10175}, 3(7):8, 2021.

\bibitem{li2021bringing}
Y Li, K Zhang, J Cao, R Timofte, and L~LocalViT Van~Gool.
\newblock Bringing locality to vision transformers.
\newblock {\em arXiv preprint arXiv:2104.05707}, 2021.

\bibitem{liang2021swinir}
Jingyun Liang, Jiezhang Cao, Guolei Sun, Kai Zhang, Luc Van~Gool, and Radu
  Timofte.
\newblock Swinir: Image restoration using swin transformer.
\newblock In {\em International Conference on Computer Vision (ICCV)}, pages
  1833--1844. IEEE, 2021.

\bibitem{lim2017enhanced}
Bee Lim, Sanghyun Son, Heewon Kim, Seungjun Nah, and Kyoung Mu~Lee.
\newblock Enhanced deep residual networks for single image super-resolution.
\newblock In {\em Computer Vision and Pattern Recognition Workshops (CVPR-W)},
  pages 136--144. IEEE/CVF, 2017.

\bibitem{liu2021swin}
Ze Liu, Yutong Lin, Yue Cao, Han Hu, Yixuan Wei, Zheng Zhang, Stephen Lin, and
  Baining Guo.
\newblock Swin transformer: Hierarchical vision transformer using shifted
  windows.
\newblock In {\em International Conference on Computer Vision (ICCV)}, pages
  10012--10022. IEEE, 2021.

\bibitem{lu2022transformer}
Zhisheng Lu, Juncheng Li, Hong Liu, Chaoyan Huang, Linlin Zhang, and Tieyong
  Zeng.
\newblock Transformer for single image super-resolution.
\newblock In {\em Computer Vision and Pattern Recognition (CVPR)}, pages
  457--466. IEEE/CVF, 2022.

\bibitem{luo2020latticenet}
Xiaotong Luo, Yuan Xie, Yulun Zhang, Yanyun Qu, Cuihua Li, and Yun Fu.
\newblock Latticenet: Towards lightweight image super-resolution with lattice
  block.
\newblock In {\em European conference on computer vision (ECCV)}, pages
  272--289. Springer, 2020.

\bibitem{luo2022deep}
Ziwei Luo, Haibin Huang, Lei Yu, Youwei Li, Haoqiang Fan, and Shuaicheng Liu.
\newblock Deep constrained least squares for blind image super-resolution.
\newblock In {\em Proceedings of the IEEE/CVF Conference on Computer Vision and
  Pattern Recognition}, pages 17642--17652, 2022.

\bibitem{martin2001database}
David Martin, Charless Fowlkes, Doron Tal, and Jitendra Malik.
\newblock A database of human segmented natural images and its application to
  evaluating segmentation algorithms and measuring ecological statistics.
\newblock In {\em International Conference on Computer Vision (ICCV)},
  volume~2, pages 416--423. IEEE, 2001.

\bibitem{matsui2017sketch}
Yusuke Matsui, Kota Ito, Yuji Aramaki, Azuma Fujimoto, Toru Ogawa, Toshihiko
  Yamasaki, and Kiyoharu Aizawa.
\newblock Sketch-based manga retrieval using manga109 dataset.
\newblock {\em Multimedia Tools and Applications}, 76:21811--21838, 2017.

\bibitem{meng2022adavit}
Lingchen Meng, Hengduo Li, Bor-Chun Chen, Shiyi Lan, Zuxuan Wu, Yu-Gang Jiang,
  and Ser-Nam Lim.
\newblock Adavit: Adaptive vision transformers for efficient image recognition.
\newblock In {\em Computer Vision and Pattern Recognition (CVPR)}, pages
  12309--12318. IEEE/CVF, 2022.

\bibitem{niu2020single}
Ben Niu, Weilei Wen, Wenqi Ren, Xiangde Zhang, Lianping Yang, Shuzhen Wang,
  Kaihao Zhang, Xiaochun Cao, and Haifeng Shen.
\newblock Single image super-resolution via a holistic attention network.
\newblock In {\em European conference on computer vision (ECCV)}, pages
  191--207. Springer, 2020.

\bibitem{qiu2023dual}
Yajun Qiu, Qiang Zhu, Shuyuan Zhu, and Bing Zeng.
\newblock Dual circle contrastive learning-based blind image super-resolution.
\newblock {\em IEEE Transactions on Circuits and Systems for Video Technology},
  2023.

\bibitem{timofte2018challenge}
R Timofte, S Gu, J Wu, and L~NTIRE Van~Gool.
\newblock challenge on single image super-resolution: Methods and results.
\newblock In {\em Computer Vision and Pattern Recognition (CVPR)}, pages
  18--22. IEEE/CVF, 2018.

\bibitem{vaswani2017attention}
Ashish Vaswani, Noam Shazeer, Niki Parmar, Jakob Uszkoreit, Llion Jones,
  Aidan~N Gomez, {\L}ukasz Kaiser, and Illia Polosukhin.
\newblock Attention is all you need.
\newblock {\em Advances in Neural Information Processing Systems (NeurIPS)},
  30, 2017.

\bibitem{wu2021cvt}
Haiping Wu, Bin Xiao, Noel Codella, Mengchen Liu, Xiyang Dai, Lu Yuan, and Lei
  Zhang.
\newblock Cvt: Introducing convolutions to vision transformers.
\newblock In {\em International Conference on Computer Vision (ICCV)}, pages
  22--31. IEEE, 2021.

\bibitem{xiao2021early}
Tete Xiao, Mannat Singh, Eric Mintun, Trevor Darrell, Piotr Doll{\'a}r, and
  Ross Girshick.
\newblock Early convolutions help transformers see better.
\newblock {\em Advances in Neural Information Processing Systems (NeurIPS)},
  34:30392--30400, 2021.

\bibitem{yoo2023enriched}
Jinsu Yoo, Taehoon Kim, Sihaeng Lee, Seung~Hwan Kim, Honglak Lee, and Tae~Hyun
  Kim.
\newblock Enriched cnn-transformer feature aggregation networks for
  super-resolution.
\newblock In {\em Winter Conference on Applications of Computer Vision (WACV)},
  pages 4956--4965. IEEE/CVF, 2023.

\bibitem{yu2023dipnet}
Lei Yu, Xinpeng Li, Youwei Li, Ting Jiang, Qi Wu, Haoqiang Fan, and Shuaicheng
  Liu.
\newblock Dipnet: Efficiency distillation and iterative pruning for image
  super-resolution.
\newblock In {\em Proceedings of the IEEE/CVF Conference on Computer Vision and
  Pattern Recognition}, pages 1692--1701, 2023.

\bibitem{zeyde2012single}
Roman Zeyde, Michael Elad, and Matan Protter.
\newblock On single image scale-up using sparse-representations.
\newblock In {\em International Conference Curves and Surfaces}, pages
  711--730. Springer, 2012.

\bibitem{zhang2022accurate}
Jiale Zhang, Yulun Zhang, Jinjin Gu, Yongbing Zhang, Linghe Kong, and Xin Yuan.
\newblock Accurate image restoration with attention retractable transformer.
\newblock {\em arXiv preprint arXiv:2210.01427}, 2022.

\bibitem{zhang2018image}
Yulun Zhang, Kunpeng Li, Kai Li, Lichen Wang, Bineng Zhong, and Yun Fu.
\newblock Image super-resolution using very deep residual channel attention
  networks.
\newblock In {\em European conference on computer vision (ECCV)}, pages
  286--301. Springer, 2018.

\bibitem{zhang2018residual}
Yulun Zhang, Yapeng Tian, Yu Kong, Bineng Zhong, and Yun Fu.
\newblock Residual dense network for image super-resolution.
\newblock In {\em Computer Vision and Pattern Recognition (CVPR)}, pages
  2472--2481. IEEE/CVF, 2018.

\bibitem{zhu2023attention}
Qiang Zhu, Pengfei Li, and Qianhui Li.
\newblock Attention retractable frequency fusion transformer for image super
  resolution.
\newblock In {\em Proceedings of the IEEE/CVF Conference on Computer Vision and
  Pattern Recognition}, pages 1756--1763, 2023.

\end{thebibliography}
}

\newpage

\section{Supplementary}
In this supplementary material, we provide additional information about model analysis and experimental results of Composite Fusion Attention Transformer (CFAT). We discuss the performance of our architecture and its variants along with their complexity in Sec. \ref{sec:Model_Analysis}. In Sec. \ref{sec:Extensive_Ablation_Study}, we present the extensive results of an ablation study related to CFAT. We also compare the proposed model with various transformer-based state-of-the-art architectures graphically and based on LAM score \cite{gu2021interpreting} in the Sec. \ref{sec:Performance_Comparison}.

\subsection{Extensive Model Analysis}
\label{sec:Model_Analysis}
We examine how the performance of our model varies alongside changes in complexity under two different settings: (i) channel size variation and (ii) model size variation. All the performances are evaluated on the BSD100 dataset for a scale of $\times 4$. In the first setting, we evaluate the complexity and the performance by increasing the value of channel counts from $96\rightarrow 144\rightarrow180\rightarrow192$ as shown in Tab. \ref{table:table1}. We observe a significant rise in performances (both PSNR and SSIM) when we widen the channel counts from $96\rightarrow 144\rightarrow180$. However, it comes with a computational burden in terms of parameter counts and Multiply-Add operations. After 180, we observe a saturation in performance, e.g., only a 0.01dB improvement in PSNR when the channel is increased from 180 to 192.  

\begin{figure}[h]
	\centering
	\captionsetup{justification=centering}
	\includegraphics[width=1.0\linewidth]{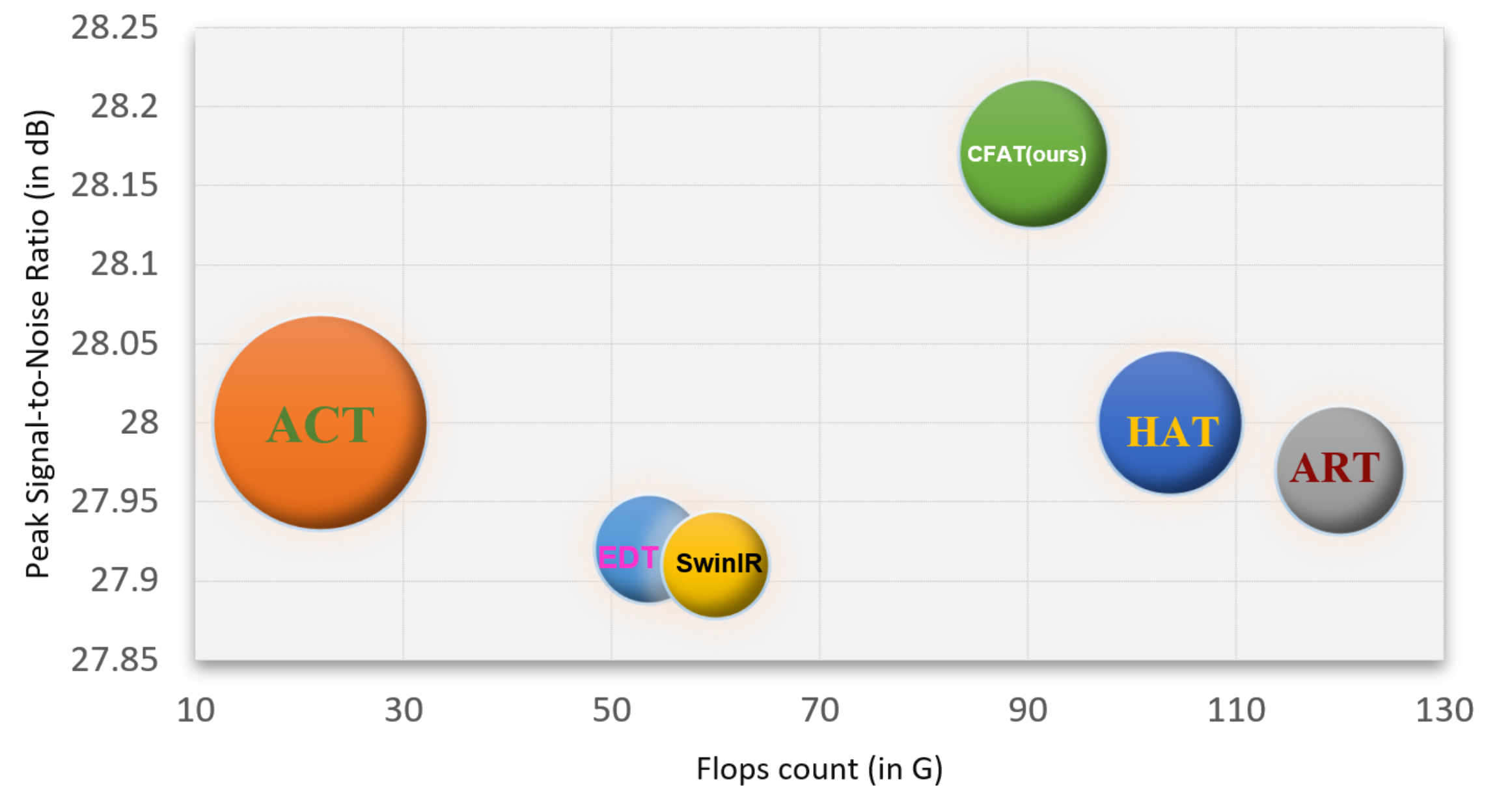}
	\vspace*{-3mm}
	\caption{Performance vs Complexity plot of CFAT compare to other state-of-the-art models. \\ \textbf{Performance:} PSNR (on X-axis) in dB. \textbf{Complexity:} Flops (on Y-axis) in G  and Parameters (area of the circle) in M }
	\label{fig:image1}
	\vspace*{-6mm}
\end{figure}

After the channel-centric evaluation of CFAT, we set the channel count to 180 in our final model and compared it with various state-of-the-art architectures of the same channel count, as shown in Fig. \ref{fig:image1}. CFAT achieves the best performance and also maintains an excellent trade-off between parameter count and number of multiply-add operations. The Multiply-Add operations in HAT \cite{chen2023activating} and ART \cite{zhang2022accurate} are too high with moderate parameter counts, whereas SwinIR \cite{liang2021swinir}, EDT \cite{li2021efficient}, and ACT \cite{yoo2023enriched} show the opposite trends. We observe an identical performance exhibited by ACT, ART, and HAT with little variations whereas the outcomes from EDT and SwinIR are comparively lesser. 

\begin{table}[h]
	\begin{center}
		\tabcolsep=0.13cm
		\singlespacing
		\vspace*{-4mm}
		\centering
		\captionsetup{justification=centering}
		\caption{Analysis of CFAT based on channel counts.}
		\vspace*{-3mm}
		\scalebox{1.0}{
			\begin{tabular}{cccc}
				\toprule
				\toprule
				Channels & Params (M) & Multi-Adds (G) & PSNR/SSIM \rule{0pt}{4ex} \\
				192 & 25.01 & 102.6 & 28.18dB/0.7524 \\
				\midrule
				180  & 22.07 & 90.59& 28.17dB/0.7524 \\
				\midrule
				144  & 14.35 & 59.22 & 27.99dB/0.7504 \\
				\midrule
				96  & 6.74  & 28.18 & 27.78dB/0.7469 \\
				\bottomrule
				\bottomrule
		\end{tabular}}
		\label{table:table1}
	\end{center}
	\vspace*{-6mm}
\end{table}

\begin{table}[h]
	\begin{center}
		\tabcolsep=0.13cm
		\singlespacing
		\vspace*{-4mm}
		\centering
		\captionsetup{justification=centering}
		\caption{Analysis of CFAT based on model size.}
		\vspace*{-3mm}
		\scalebox{1.0}{
			\begin{tabular}{cccc}
				\toprule
				\toprule
				Models & Params (M) & Multi-Adds (G) & PSNR/SSIM \rule{0pt}{4ex} \\
				\midrule
				CFAT-l  & 34.89 & 142.08 & 28.25dB/0.7531 \\
				\midrule
				CFAT  & 22.07  & 90.59 & 28.17dB/0.7524 \\
				\midrule
				CFAT-s  & 14.35 & 59.22 & 27.99dB/0.7504 \\
				\midrule
				CFAT-r  & 13.52 & 56.27 & 27.93dB/0.7498 \\
				\bottomrule
				\bottomrule
		\end{tabular}}
		\label{table:table2}
	\end{center}
	\vspace*{-6mm}
\end{table}

\begin{figure*}[h]
	\centering
	\captionsetup{justification=centering}
	\includegraphics[width=0.825\linewidth]{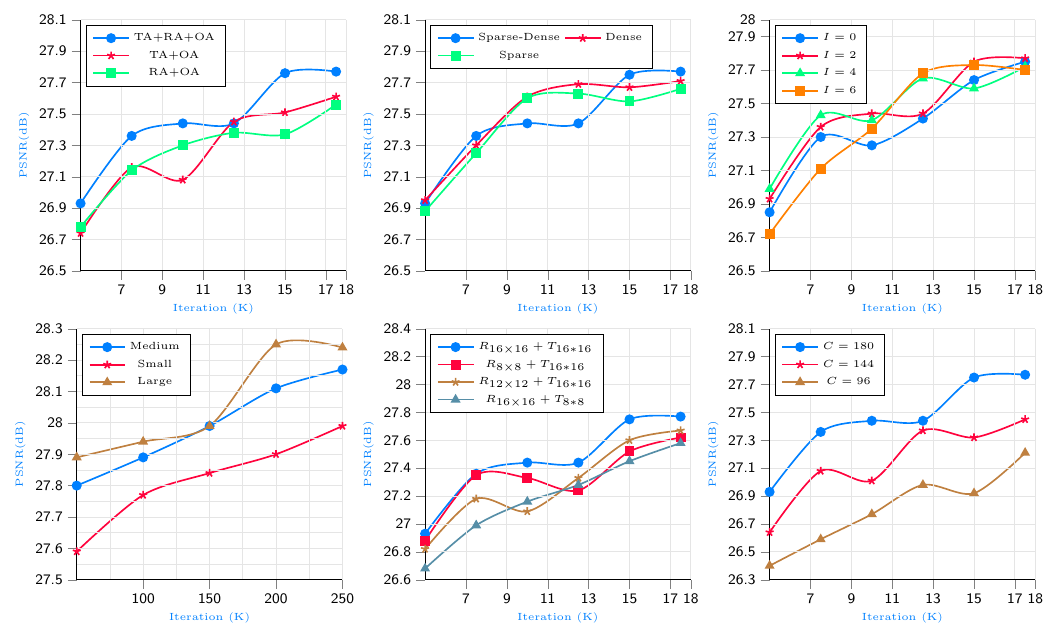}
	\vspace*{-4mm}
	\caption{Iterative performance (PSNR in dB) comparison of the proposed CFAT for \textbf{Top-Left:} triangular vs rectangular vs overlapping attention, \textbf{Top-Middle:} sparse vs dense attention, \textbf{Top-Right:} various interval size, \textbf{Bottom-Left:} small vs medium vs large CFAT model, \textbf{Bottom-Middle:} various combinations of rectangular ($8\times8$, $12\times12$, $16\times16$) with triangular ($8*8$, $16*16$) windows, and \textbf{Bottom-Right:} various channel lengths. [on BSD100($\times4$) for epoch 70]}
	\vspace*{-6mm}
	\label{fig:image_2}
\end{figure*}

\begin{figure*}[h]
	\centering
	\captionsetup{justification=centering}
	\includegraphics[width=0.825\textwidth,height=0.5\textheight]{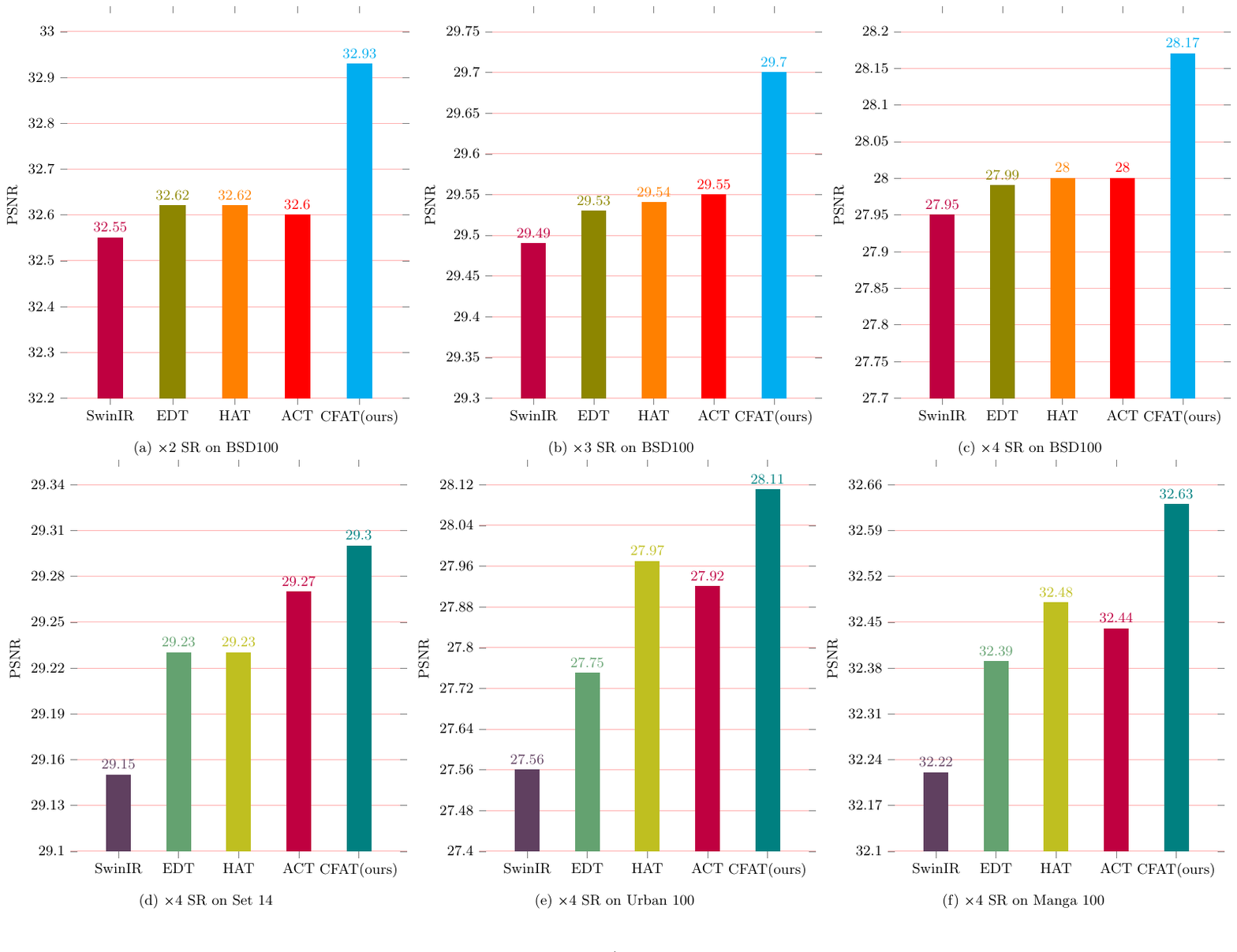}
	\vspace*{-8mm}
	\caption{Comparing performance (PSNR in dB) of various state-of-the-art models with CFAT on \textbf{Top-Left:} BSD100 for scale 2, \textbf{Top-Middle:} BSD100 for scale 3, \textbf{Top-Right:} BSD100 for scale 4, \textbf{Bottom-Left:} Set14 for scale 4,\\ \textbf{Bottom-Middle:} Urban100 for scale 4, and \textbf{Bottom-Right:} Manga109 for scale 4.}
	\label{fig:image_3}
\end{figure*}

We consider Dense Window Attention Blocks (DWAB) and Sparse Window Attention Blocks (SWAB) to be the basic units of CFAT and termed Window Attention Blocks (WAB). We compose three architectures, CFAT-l (large), CFAT (medium), and CFAT-r (reduced), based on model depth, i.e., the number of WAB units present in CFAT. We take (8, 8, 8, 8, 8, 8, 8, 8) WAB units for CFAT-l, (8, 8, 8, 8, 8) for CFAT, and (8, 8, 8) for CFAT-r. Here, '8' signifies '4' pairs of Shifted-Dense Rectangular Window MSA ((SD)RW-MSA) and Shifted-Dense Triangular Window MSA ((SD)TW-MSA) arranged in an alternative fashion. We also consider another CFAT variant with 144 channels to compete with CFAT-r while finalizing our small variant. The corresponding parameters, Multiply-Add operations, and performances are displayed in Tab. \ref{table:table2}. To finalize the small version, we prioritize more on performance over complexity. This table shows that CFAT with 144 channels yields higher performance than CFAT-r, while both possess identical complexity levels. Therefore, we designate CFAT with 144 channels as our small variant, CFAT-s.
\begin{figure*}[t]
	\centering
	\vspace*{-1mm}
	\captionsetup{justification=centering}
	\includegraphics[width=0.8\linewidth]{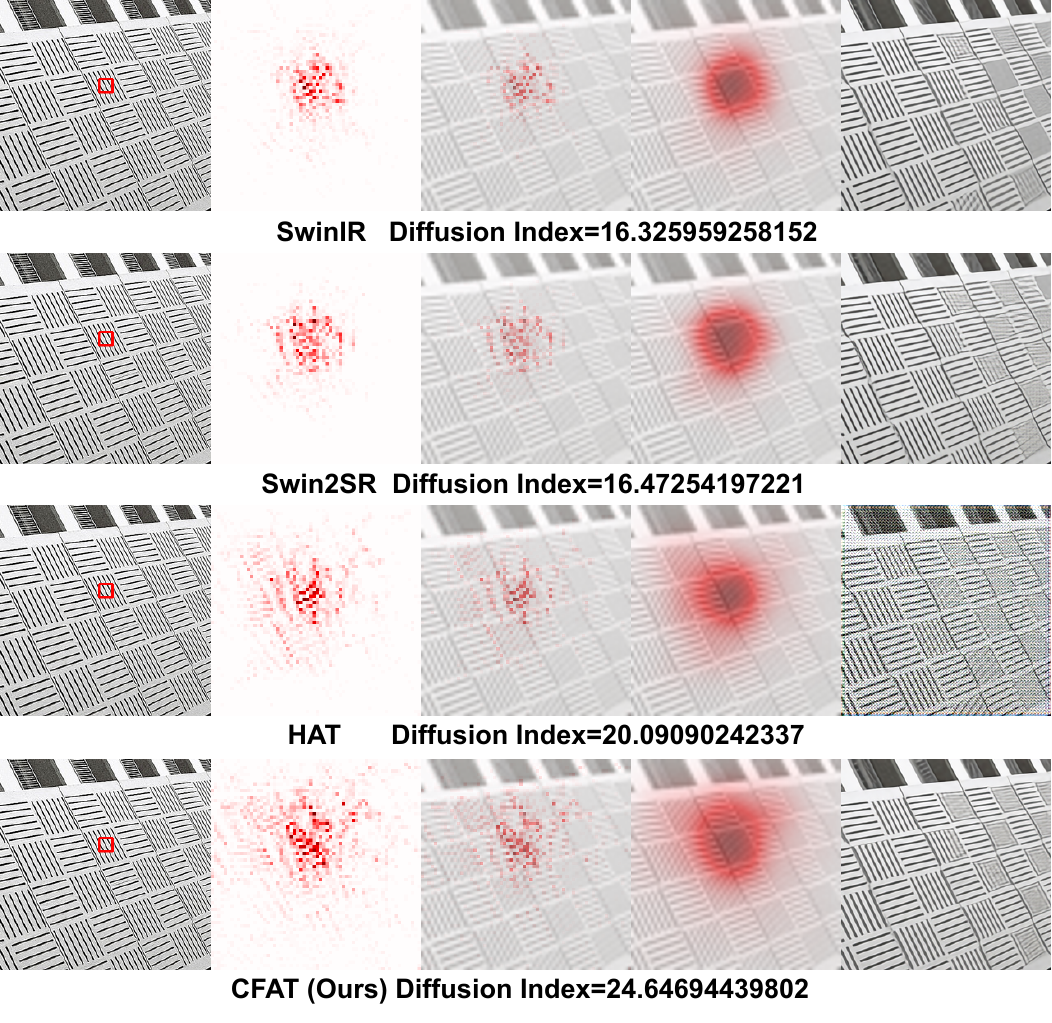}
	\vspace*{-4mm}
	\caption{LAM results and corresponding Diffusion Index for CFAT and various SOTA methods.}
	\label{fig:LAM}
	\vspace*{-5.5mm}
\end{figure*}

\subsection{Extensive Ablation Study}
\label{sec:Extensive_Ablation_Study}
In this section, we investigate the performance variation of CFAT under the influence of different hyperparameters and model units. We evaluate all the model variants on the BSD100 dataset for scale factor $\times4$ at distinct training iterations. We plot these outcomes in the X-axis along with their respective iterations in the Y-axis as displayed in Fig. \ref{fig:image_2}. A large deviation in performance is observed when we evaluate the model within the first 17.5k iterations. Therefore, we adopt an averaging technique to keep the results steady at iterations of 5k, 7.5k, 10k, 12.5k, 15k, and 17.5k. The averaging technique is implemented within the iteration or epoch range of $\pm2.5k \text{ iteration}$ or $\pm5 \text{ epoch}$, respectively. The Top-Left plot shows the outcomes for models taking rectangular with overlapping attention, triangular with overlapping attention and rectangular, triangular with overlapping attention. We find that the last configuration yields the best results. The Top-Middle plot displays the significance of the combined dense-sparse attention-based model over isolated attention-based models. The Top-Right plot justifies selecting the interval size as '2' over others. The Bottom-Left plot maps out the performance of three CFAT-variants: CFAT-l, CFAT, and CFAT-s. Based on performances in the Bottom-Middle plot, we decide the best combination of window sizes for rectangular- and triangular-window attention. We map the model outcomes for three-channel counts (180, 144, and 96) in the Bottom-Right plot.

\vspace*{-2mm}

\subsection{Extended Comparison with SOTA Architectures}
\label{sec:Performance_Comparison}
\vspace*{-2mm}
In this section, we evaluate and compare the performance of the proposed architecture with other transformer-based state-of-the-art models. The top three graphs (Top-Left, Top-Middle, and Top-Right) of Fig. \ref{fig:image_3} validate the supremacy of CFAT over SwinIR \cite{liang2021swinir}, EDT \cite{li2021efficient}, ACT \cite{yoo2023enriched}, and HAT \cite{chen2023activating} super-resolution (SR) architectures on BSD100 testing dataset for scales of $\times2$, $\times3$, and $\times4$. These graphs also justify that our model possesses strong expressive power for every scale of super-resolution. As displayed in the bottom three graphs (Bottom-Left, Bottom-Middle, and Bottom-Right), we also verify the generalizability of CFAT by evaluating the performance on different testing datasets for a fixed scale ($\times4$). All these performances are expressed in terms of peak-signal-to-noise ratio (PSNR). CFAT yields the highest PSNR values for all the above settings, as shown in this figure.

As visualized in Fig. \ref{fig:LAM}, the LAM attributes \cite{gu2021interpreting} and Diffusion Index (DI) \cite{gu2021interpreting} of the proposed triangular window-based CFAT yield superior results than other rectangular window-based SOTA methods. To check the model's scalability in a low-data environment, all models are trained on the DIV2K dataset with batch size 16.

\end{document}